\documentclass[a4paper,11pt]{article}
\pdfoutput=1
\usepackage{jheppub}
\usepackage[T1]{fontenc}
\usepackage{amsfonts}
\usepackage{mathrsfs}
\usepackage{units}
\usepackage{url}

\newcommand{\al}{\alpha}
\newcommand{\be}{\beta}
\newcommand{\te}{\theta}
\newcommand{\tb}{t_{\beta}}

\begin{document}

\title{Can the 125 GeV Higgs be the \emph{Little Higgs}?}

\author{J. Reuter}
\author{and M. Tonini}
\affiliation{DESY Theory Group\\Notkestr. 85, 22603 Hamburg, Germany}

\emailAdd{juergen.reuter@desy.de}
\emailAdd{marco.tonini@desy.de}

\preprint{DESY 12-177}

\abstract{%
After the discovery of the Higgs-like boson by the LHC 2012
it is the most important task to check whether this new particle is the Standard Model
Higgs boson or something else. In this paper, we study whether the 125 GeV boson
could be the pseudo-Goldstone boson of Little Higgs models. We derive limits on the
parameter space of several Little Higgs models (simple group and product group models, 
with and without $T$-parity), both from the experimental data from ATLAS and CMS about 
the different Higgs discovery channel and the electroweak precision observables. 
We perform a fit of several Little Higgs models to all electroweak parameters from measurements
of SLC, LEP, Tevatron, and LHC. For the Higgs searches, we include all
available data from the summer conferences in 2012 as well as the updates
from December 2012. We show that there always exists a region in the
parameter space of the models under consideration where the measured
$\chi^{2}$ is equal or lower than the SM $\chi^{2}$: a closer look at
the minimum $\chi^{2}$ will however reveal that the agreement with the
collected data is not significantly better as within the SM. While for
the models without $T$-parity the Little Higgs scale $f$ is forced to
be of the order 2-4 TeV in order to be compatible with the collected
data, in the models with $T$-parity the scale $f$ is constrained to be
only above $\mathcal{O}(500)$ GeV, reducing the amount of
fine-tuning. We also show that these results are still driven by the
electroweak precision measurements due to the bigger LHC data
uncertainties. 
}

\maketitle

\section{Introduction}

The discovery of a bosonic particle with a mass of 125 GeV by the LHC experiments 
2012~\cite{:2012gu,:2012gk}  seems to be the last piece of the jigsaw puzzle of the 
electroweak interactions. However, at present it is not yet clear whether this particle 
does really have all the properties of the Standard Model (SM) Higgs boson, or whether 
it is a particle of some extension of the SM. 

There are many reasons to believe in the existence of beyond the Standard Model (BSM)
physics: the missing CP violation needed for the explanation of the baryon-antibaryon
asymmetry in the universe, the missing dark matter component in the SM, and the 
question about the stability of the Higgs potential and the electroweak vacuum: the latter
has been called the hierarchy or fine-tuning problem, namely the problem that the 
Higgs self-coupling is driven to non-perturbative values for too large Higgs masses
while the top couplings tend to destabilize the electroweak vacuum. Furthermore, the 
bare Higgs mass parameter seems to be tuned very accurately in order to get a Higgs
mass at the electroweak scale, as scalar masses are quadratically sensitive to new physics
particles coupling to them. 

One paradigm to solve this problem is to assume the Higgs boson to be no fundamental, but
a composite particle, as e.g. in Technicolor, Topcolor or composite Higgs models. The
Higgs boson is relatively light compared to high scales, because it appears, like the pions
in chiral symmetry breaking, as the (pseudo)-Nambu-Goldstone bosons (pNGBs) of a 
spontaneously broken global symmetry. However, this necessitates the presence of strong
interactions to bind new constituents together to something like the Higgs boson, and
indications of such strong interactions had to show up in the electroweak precision 
measurements from SLC and LEP (and also Tevatron). A solution to this problem has been found
using the formalism of collective symmetry breaking, where several global symmetries
are intertwined. If each of them were exact, the Higgs would still be an exact massless
Goldstone boson. Hence, the mass term arises only logarithmically at the one-loop order
or quadratically at two-loop order. This leaves such models weakly interacting at the TeV scale
and raises the scale for the onset of new strong interactions to several TeV up to tens of TeV.
These kinds of models have first been realized motivated from deconstructed extra 
dimensions~\cite{ArkaniHamed:2001nc,ArkaniHamed:2002pa}, and then in a 4D setup 
by explicit constructions of coset spaces for the symmetry breaking pattern~\cite{ArkaniHamed:2002qy,Low:2002ws,Kaplan:2003uc}. There are two different types
of models, so-called Simple Group Models, where the weak gauge group extension is given
by a simple Lie group, while the Goldstone multiplet of the broken global symmetry is distributed
over several different non-linear sigma model fields, whereas in the Product Group Models the
Goldstone multiplet is a single representation parameterizing the coset space of the global
symmetry breaking, and the weak gauge group emerges as the unbroken part of a product gauge
group. The most prominent examples of these two classes are the Simplest Little Higgs 
model~\cite{Schmaltz:2004de} and the Littlest Higgs model~\cite{ArkaniHamed:2002qy}, respectively. 
To ameliorate the amount of fine tuning within the so-called Little Hierarchy problem between the 
electroweak and the TeV scale, a discrete symmetry named $T$-parity has been 
introduced~\cite{Cheng:2003ju}: it cancels tree-level contributions from heavy Little Higgs states 
to the electroweak precision observables (at least in the gauge and scalar sector for the product-group
models), and offers a possibility for a dark matter particle. For an overview over and more details 
about Little Higgs models, cf.~\cite{Schmaltz:2005ky,Perelstein:2005ka}.

In this paper, we discuss the most common Little Higgs models, namely the Littlest Higgs with and
without $T$-parity as well as the Simplest Little Higgs, and fit the results reported by both experimental
collaborations, ATLAS and CMS, about the many different Higgs search/discovery channels to 
these different models. This is accompanied by a simultaneous fit of the electroweak precision data to
these models. We compare the constraints coming from the LHC Higgs discovery with those from
electroweak precision physics. The paper is organized as follows: Sec.~\ref{sec:frame} gives a 
technical introduction into the three different Little Higgs models
under consideration, structured according 
to their gauge and scalar sector, the fermion sector, and finally
discussing the electroweak precision 
observables for these models.  The experimental data needed for the analysis presented here are
given in Sec.~\ref{sec:stat_exp}, where we also present the statistical methods that we used to perform
the fit of the Little Higgs models to the experimental data as well as to the precision observables.
Our results are presented in Sec.~\ref{sec:results}, before we give our conclusions in 
Sec.~\ref{sec:conclusions} . In the appendix, technical details on the determination of the Higgs boson 
partial widths and cross section as well as on the calculation of the electroweak precision observables
within the Little Higgs models are shown.


\section{The Little Higgs framework}
\label{sec:frame}

In this section, we will describe the structure of the three different Little
Higgs models under consideration, focusing in particular on the details
which will affect our results. However, this section should not be
thought as a comprehensive review of these models, for which we refer
to~\cite{Schmaltz:2005ky,Perelstein:2005ka}. 

We decided to present separately the structure of the gauge, the
scalar and the fermion sectors, in order to underline the different
implementations of the Little Higgs paradigm in the considered
models. In the end, there is also a subsection describing the effect
of the Little Higgs structure on the predictions of Electroweak
Precision Observables (EWPO). 


\subsection{Gauge and Scalar sectors}

\subsubsection*{Littlest Higgs Model}

The Littlest Higgs model (we mainly follow the presentation given in
\cite{Han:2003wu} instead of the original
paper~\cite{ArkaniHamed:2002qy}; in the sequel, we use the
abbreviation $L^2H$) is based on a
non-linear sigma model in the coset space  
\begin{equation}
	SU(5)/SO(5).
	\label{LHT}
\end{equation}
The vacuum expectation value (\emph{vev}) of an $SU(5)$ symmetric
tensor field generates the global spontaneous symmetry breaking
$\left( \ref{LHT} \right)$ at the scale \emph{f}: 
\begin{equation}
	\langle \Sigma \rangle = \left( \begin{array}{ccc} \mathbf{0}_{2\times2} & \mathbf{0}_{2\times1}&\mathbf{1}_{2} \\ \mathbf{0}_{1\times2} & 1 & \mathbf{0}_{1\times2} \\ \mathbf{1}_{2} & \mathbf{0}_{2\times1} & \mathbf{0}_{2\times2}  \end{array} \right)
\end{equation}
In this setup, there are 14 Nambu-Goldstone Bosons (NGBs) $\Pi^{a}$, $a=1,\ldots,14$, parametrized by 
\begin{equation}
	\Sigma(x)=e^{2 \, i \, \Pi^{a} X^{a}(x)/f} \, \langle \Sigma \rangle
\end{equation}
where $X^{a}$ are the broken generators of the coset space $SU(5)/SO(5)$. 

This model belongs to the class of Product Group models, where the SM
gauge group emerges from the diagonal breaking of the product of
several gauged groups: in this specific realization there is a local
invariance under $[SU(2)_{1} \otimes U(1)_{1}] \otimes [SU(2)_{2}
\otimes U(1)_{2}]$, embedded in the matrix structure, spontaneously
broken through the \emph{vev} $\langle \Sigma \rangle$ to its diagonal
subgroup, which is identified with the SM gauge group. A set of
$SU(2) \otimes U(1)$ gauge bosons obtains a mass of order
\emph{f}, while the other set is left massless and is identified with
the SM gauge fields.  

Under the unbroken $SU(2)_{L} \otimes U(1)_{Y}$ the $\Pi^{a}$
transform as $\textbf{1}_{0} \oplus \textbf{3}_{0} \oplus
\textbf{2}_{1/2} \oplus \textbf{3}_{\pm 1}$: the $\textbf{2}_{1/2}$
component is identified with the Higgs boson \emph{h}, while the
$\textbf{3}_{\pm 1}$ component is a complex triplet under $SU(2)_{L}$
which forms a symmetric tensor $\Phi_{ij}\equiv \Phi$ with components
$\phi^{++}$, $\phi^{+}$, $\phi^{0}$ and a pseudo-scalar $\phi^{P}$,
where both $\phi^{0}$ and $\phi^{P}$ are real scalars. The other
components are the longitudinal modes of the heavy gauge bosons and therefore will not appear in unitary gauge.

The kinetic term for the NGB matrix can be expressed in the standard
non-linear sigma model formalism as 
\begin{equation}
	\mathcal{L}_{\Sigma} = \frac{1}{2} \frac{f^2}{4} \text{ tr}
        \big| D_{\mu} \Sigma \big|^2 
	\label{scalkin}
\end{equation}
where the numerical coefficients assure canonically normalized kinetic
terms for the scalar fields. To impose a local invariance under
$[SU(2)_{1} \otimes U(1)_{1}] \otimes [SU(2)_{2} \otimes U(1)_{2}]$,
the covariant derivative is defined as 
\begin{equation}
	D_{\mu} \Sigma = \partial_{\mu} \Sigma - i \sum_{j=1}^{2}
        \left[ g_j (W_j \Sigma + \Sigma W_j^t)+g_j^{\prime} (B_j
          \Sigma + \Sigma B_j^t) \right] 
\end{equation}
and the generators of the gauged symmetries are explicitly given as 
\begin{eqnarray}
	Q_{1}^{a} &=& \left( \begin{array}{ccc} \sigma^{a}/2 & 0 & 0
            \\ 0 & 0 & 0 \\ 0 & 0 & 0 \end{array} \right) \qquad \quad
        Y_{1} = \text{diag} \left( 3,3,-2,-2,-2 \right)/10 \nonumber
        \\ 
	Q_{2}^{a} &=& \left( \begin{array}{ccc} 0 & 0 & 0 \\ 0 & 0 & 0
            \\ 0 & 0 & -\sigma^{a *}/2 \end{array}  \right) \qquad
        Y_{2} = \text{diag} \left( 2,2,2,-3,-3 \right)/10. 
\end{eqnarray}

The global symmetries prevent the appearance of a potential for the
scalar fields at tree level. The scalar potential is indeed generated
at one-loop and higher orders due to the interactions with gauge
bosons and fermions, and is parametrized through the Coleman-Weinberg
(CW) potential \cite{Coleman:1973jx}. The scalar potential takes the
generic form 
\begin{equation}
	V_{CW} = \lambda_{\phi^2} f^2 \text{ tr}(\phi^{\dagger} \phi)
        + i \lambda_{h \phi h} f (h \phi^{\dagger} h^T-h^* \phi
        h^{\dagger})-\mu^2 h h^{\dagger}+ \lambda_{h^4} (h
        h^{\dagger})^2 
\end{equation}
where the coefficients $\lambda_{\phi^2}$, $\lambda_{h \phi h}$ and
$\lambda_{h^4}$ are functions of the fundamental parameters of the
model, while the Higgs mass parameter $\mu^2$ should be treated as a
free parameter since it receives big contributions also from two-loop
diagrams, that have not been calculated.  

Minimizing the potential to obtain the doublet and triplet \emph{vev}s
$v$ and $v^{\prime}$, and requiring appropriate relations to correctly
trigger electroweak symmetry breaking (EWSB), one can express all four
parameters in the scalar potential to leading order in terms of the
physical parameters $f$, $m_h^2$, $v$ and $v^{\prime}$, and obtain the
following relation between the two \emph{vev}s 
\begin{equation}
	x \equiv \frac{4 v^{\prime} f}{v^2}, \qquad 0 \leq x < 1 \; .
	\label{xdef}
\end{equation}
Diagonalizing the scalar mass matrix, one obtains at leading order the following spectrum:
\begin{equation}
	m_h = \sqrt{2} \mu, \qquad m_{\Phi} = \frac{\sqrt{2}
          m_h}{\sqrt{1- x^2}} \frac{f}{v} \qquad ,
\end{equation}
where all components of the triplet $\left( \phi^{++}, \phi^{+},
  \phi^{0}, \phi^{P} \right)$ are degenerate at the order we are
considering. Since $\mu^2$ is treated as a free parameter, we will
assume the measured Higgs mass for the scalar doublet $h$, fixing
therefore the value of $\mu$.  

If we parametrize the interaction terms of the charged components of
the triplet to the Higgs field in the following way 
\begin{equation}
	V_{CW} \supset -2 \, \frac{m_{\Phi}^{2}}{v} \, y_{\phi^{+}} \,
        \phi^{+} \phi^{-} h - 2 \, \frac{m_{\Phi}^{2}}{v} \,
        y_{\phi^{++}} \, \phi^{++} \phi^{--} h 
\end{equation}
then the couplings $y_{\phi}$ after EWSB are predicted up to
$\mathcal{O} \left(v^{2}/f^{2} \right)$ to be~\cite{Wang:2011rv} 
\begin{equation}
	y_{\phi^{+}} = \frac{v^2}{f^2} \left( -\frac{1}{3} +
          \frac{x^2}{4} \right), \qquad y_{\phi^{++}} = \mathcal{O}
        \left( \frac{v^{4}}{f^{4}} \right). 
\end{equation}
This parametrization will be useful to calculate the contribution of
the charged resonances to the one-loop $h \gamma \gamma$ vertex,
cf. Appendix~\ref{app:higgs}. 

A set of $SU(2) \otimes U(1)$ gauge bosons ($W^{\prime}$,
$B^{\prime}$) obtains a mass term of order $f$ from ($\ref{scalkin}$),
while the other set ($W$, $B$) remains massless. The mass eigenstates
are related to the gauge eigenstates by the following field rotations 
\begin{eqnarray}
	W &=& s W_1 + c W_2, \qquad W^{\prime} = -c W_1+s W_2 \\ 
	B &=& s^{\prime} B_1 + c^{\prime} B_2 \qquad B^{\prime} =
        -c^{\prime} B_1 + s^{\prime} B_2 \nonumber 
\end{eqnarray}
where the mixing angles, which we will treat as free parameters, are given by
\begin{equation}
	c = \frac{g_1}{\sqrt{g_1^2+g_2^2}}, \qquad c^{\prime} =
        \frac{g^{\prime}_1}{\sqrt{g_1^{\prime \, 2}+g_2^{\prime \,
              2}}}. 
	\label{cdef}
\end{equation}

EWSB induces further mixing between the light and heavy gauge bosons:
at leading order, the spectrum is given by 
\begin{eqnarray}
	m_{W^{\pm}} = \frac{g v}{2} & \qquad & m_{W_{H}^{\pm}} =
        \frac{g f}{2 s c} \nonumber \\ 
	m_{Z} = \frac{g v}{2 c_w} & \qquad & m_{Z_{H}} = \frac{g f}{2
          s c} \\ 
	m_{\gamma}=0 & \qquad & m_{A_{H}} = \frac{g^{\prime} f}{2
          \sqrt{5} s^{\prime} c^{\prime}} \nonumber 
\end{eqnarray}
If we parametrize the interaction terms of the charged gauge bosons to
the Higgs field as 
\begin{equation}
	\mathcal{L}_{\Sigma} \supset 2 \, \frac{m_{W}^{2}}{v} \, y_{W} \, W^{+} W^{-} h + 2 \, \frac{m_{W_{H}}^{2}}{v} \, y_{W_{H}} \, W_{H}^{+} W_{H}^{-} h
\end{equation}
then the couplings $y_{V}$ after EWSB are predicted up to $\mathcal{O}
\left(v^{2}/f^{2} \right)$ to be \cite{Wang:2011rv} 
\begin{equation}
	y_{W} = 1 + \frac{v^2}{f^2} \left[ -\frac{1}{6} - \frac{1}{4}
          (c^2-s^2)^2 \right], \qquad y_{W_{H}} = -s^2 c^2
        \frac{v^2}{f^2}. 
\end{equation}

The $L^2H$ model contains new matter content and interactions which
contribute to the EWPO, as we will discuss in detail later. In particular, from
the exchange of heavy $SU(2)$ gauge bosons and from the presence of
the triplet \emph{vev} $v^{\prime}$, the relation between the Fermi
constant $G_F$ and the doublet \emph{vev} $v$ is modified from its SM
form: by comparing the two relations one can thus express the $L^2H$
doublet \emph{vev} $v$ in terms of the SM value $v_{SM}=246$ GeV up to
$\mathcal{O} \left( v_{SM}^2/f^2 \right)$ as \cite{Wang:2011rv} 
\begin{equation}
	v = v_{SM} \left[ 1- \frac{v_{SM}^2}{f^2} \left( -\frac{5}{24} + \frac{x^2}{8} \right) \right].
	\label{vevLH}
\end{equation}
Using this relation, we can express the corrections of the SM-like
$hVV$ couplings (V $\equiv$ W, Z) with respect to their SM values up
to $\mathcal{O} \left( v_{SM}^2/f^2 \right)$ equivalently as 
\begin{equation}
	\frac{g_{hVV}}{g_{hVV}^{SM}} = 1 +
        \frac{1}{8}\frac{v_{SM}^2}{f^2} \bigg[ -3 +x^2 -2(c^2-s^2)^2
        \bigg] 
	\label{hVVLH}
\end{equation}
where 
\begin{equation}
	g_{hVV}= \frac{m_{V}^{2}}{v} \, y_{V}, \qquad g_{hVV}^{SM} = \frac{m_{V}^2}{v}_{\big{|_{v=v_{SM}}}}.
\end{equation}
Eq. ($\ref{hVVLH}$) will be useful to calculate the tree-level decays
of the Higgs boson into the SM-like gauge bosons, cf. Appendix~\ref{app:higgs} .

We will not consider all other tree-level decay channels of the Higgs
which involve the heavy gauge bosons or the heavy scalar triplet:
indeed in $L^2H$ the EWPD require $f$ larger than a few TeV,
cf.~Ref.~\cite{Csaki:2002qg} and  our results of
Sec.~$\ref{sec:results}$, making these decay channels kinematically
forbidden.  


\subsubsection*{Littlest Higgs with \emph{T}-parity}

As just mentioned, the original Littlest Higgs model suffers from
severe constraints from EWPO, which could only be satisfied in
small regions of the parameter space. The most severe constraints
resulted from tree-level corrections to EWPO due to the
exchange of the heavy gauge bosons present in the theory, as well as
from the small but non-vanishing \emph{vev} of the additional scalar
triplet field $\Phi$. These severe constraints are evaded with the
introduction of a conserved discrete symmetry, called \emph{T-parity},
featuring \emph{T}-odd partners for all (\emph{T}-even) SM particles,
and a lightest \emph{T}-odd particle that is stable. As a result,
tree-level contributions of the heavy gauge bosons to EWPO are
suppressed, and corrections arise only at loop level. 

The Littlest Higgs model with \emph{T}-parity (for detailed
reviews cf.~\cite{Hubisz:2004ft,Hubisz:2005tx}, and the original
papers~\cite{Cheng:2003ju,Cheng:2004yc}; in the following we use the
abbreviation \emph{LHT}) shares the same global and local symmetry
structure of the original $L^2H$ model. The \emph{LHT} model has
therefore the same scalar kinetic term of Eq. ($\ref{scalkin}$), where
the \emph{T}-parity can be naturally implemented requiring that the
coupling constant of $SU(2)_1$ ($U(1)_1$) equals that of $SU(2)_2$
($U(1)_2$): in this way the four mixing angles of the gauge sector
$c,s,c^{\prime},s^{\prime}$ are all equal to $1/\sqrt{2}$.  

Under \emph{T}-parity, the Higgs field and the SM-like gauge bosons
are \emph{T}-even, while the scalar triplet and the heavy gauge bosons
are \emph{T}-odd. Therefore the coupling $h^{\dagger} \Phi h$ is
forbidden, leading to the relations for the triplet \emph{vev} $v^{\prime}=0$ and
$x=0$. Since the correction of $W_H$ to the relation between $G_F$ and
$v$ is forbidden by \emph{T}-parity, the functional form of the Higgs
\emph{vev} $v$ up to $\mathcal{O} \left( v_{SM}^2/f^2 \right)$ is
modified as \cite{Wang:2011rv} 
\begin{equation}
	v = v_{SM} \left( 1+\frac{1}{12} \frac{v_{SM}^{2}}{f^{2}} \right).
	\label{vevLHT}
\end{equation}

The scalar and gauge boson mass spectrum and their couplings to the
Higgs field in \emph{LHT} model can be easily obtained from the
respective $L^2H$ relations by taking
$c=s=c^{\prime}=s^{\prime}=1/\sqrt{2}$ and $x=0$. Only the $hVV$
coupling ($V \equiv W, Z$) gets a different correction in the
\emph{LHT} model because of the different functional form of $v$
\cite{Chen:2006cs}: 
\begin{equation}
	\frac{g_{hVV}}{g_{hVV}^{SM}} = 1- \frac{1}{4} \frac{v_{SM}^{2}}{f^{2}}-\frac{1}{32} \frac{v_{SM}^{4}}{f^{4}} + \mathcal{O} \left( \frac{v_{SM}^6}{f^6} \right).
	\label{hVVLHT}
\end{equation}

If the lightest \emph{T}-odd particle $A_H$ is very light, also the
tree-level decay $h \rightarrow A_H A_H$ could be kinematically open
in \emph{LHT}. The $hA_{H}A_{H}$ coupling is given by
\cite{Han:2003wu} 
\begin{equation}
	g_{h A_{H} A_{H}} = -\frac{1}{2} \, g^{\prime \, 2} \, v \qquad,
\end{equation}
and in Appendix~\ref{app:higgs} there is the explicit expression of
the partial width of this decay channel: indeed in \emph{LHT} a lower
value of $f$ is allowed by EWPO \cite{Hubisz:2005tx}, and thus
this decay channel could be kinematically open. Note that if one
assumes the $A_H$ to be the dark matter particle, a resonant
coannihilation of two heavy photons via $s$-channel Higgs exchange is
actually favored, rendering this channel close to irrelevant. 


\subsubsection*{Simplest Little Higgs}

The Simplest Little Higgs (for detailed reviews
cf.~\cite{Han:2005ru,Cheung:2006nk,delAguila:2011wk}, while the
original references are~\cite{Kaplan:2003uc,Schmaltz:2004de}, and
the used abbreviation \emph{SLH})
is based on a non-linear sigma model in the coset space  
\begin{equation}
	\frac{\left[ SU(3)_1 \otimes U(1)_1 \right] \otimes \left[
            SU(3)_2 \otimes U(1)_2 \right]}{\left[ SU(2)_1 \otimes
            U(1)_1 \right] \otimes \left[ SU(2)_2 \otimes U(1)_2
          \right]}. 
	\label{SLH}
\end{equation} 
The \emph{vev}s of two $SU(3)_1 \otimes SU(3)_2$ scalar fields
$\phi_{1} \sim (\mathbf{3},\mathbf{1})$ and $\phi_2 \sim
(\mathbf{1},\mathbf{3})$ realize the spontaneous symmetry breaking
$SU(3)_i \rightarrow SU(2)_i$ $(i=1,2)$ at scales $f_{1}$ and $f_{2}$
respectively, giving rise to ten NGBs.  

This model belongs to the class of Simple Group models, where the SM
gauge group emerges from the breaking of a larger simple group: in
this specific realization there is a local invariance under the
diagonal subgroup $SU(3)_L \otimes U(1)_X$, which is spontaneously
broken by the \emph{vev}s of $\phi_{1,2}$ to the SM $SU(2)_{L} \otimes
U(1)_{Y}$. Five NGBs are therefore eaten and five new gauge bosons
arise with a mass of the order of the scale $f$, with
$f^{2}=f_{1}^{2}+f_{2}^{2}$. 

The NGBs are parametrized with a non-linear representation of the two
complex scalar triplet fields $\phi_{1,2}$ 
\begin{equation}
	\phi_{1}(x) = \exp{\left( \frac{i \tb \Theta(x)}{f} \right)}
        \left( \begin{array}{c} 0 \\ 0 \\ f c_{\be} \end{array}
        \right), \qquad \phi_{2}(x) = \exp{\left( -\frac{i
              \Theta(x)}{\tb f} \right)} \left( \begin{array}{c} 0 \\
            0 \\ f s_{\be} \end{array} \right) 
	\label{scalarsSLH}
\end{equation}
with $\tb=\sin{\be}/\cos{\be}=f_2/f_1$ being the ratio of the \emph{vev}s of
the scalar triplets, and $\Theta(x)$ the NGB matrix 
\begin{equation}
	\Theta = \frac{1}{f} \left[ \left( \begin{array}{cc}
              \mathbf{0}_{2\times2} & h \\ h^{\dagger} & 0 \end{array}
          \right) + \frac{\eta}{\sqrt{2}} \mathbf{1}_{3\times3}
        \right] .
\end{equation}
Here, we have already neglected (in unitary gauge) the NGBs that
become the longitudinal modes of the 5 heavy and 3 SM-like gauge
bosons (after EWSB): indeed the remaining 2 physical NGBs are
identified with the Higgs doublet $h$ and with a pseudo-scalar $\eta$
as above. 

The presence of the pseudo-scalar $\eta$ and in particular of the
coupling $h$-$Z$-$\eta$ is a peculiar and distinguishing feature of
Simple Group models class, as already pointed out in
\cite{Kilian:2006eh}. 

The kinetic term for the scalar sector can be expressed in the standard non-linear sigma model formalism as
\begin{equation}
	\mathcal{L}_{\Phi} = \sum_{i=1}^{2} \big| D_{\mu} \phi_i \big|^2
	\label{scalkinSLH}
\end{equation}
where the covariant derivative, in order to assure $SU(3)_L \otimes U(1)_X$ local invariance, is
\begin{equation}
	D_{\mu} = \partial_{\mu} - i g A_{\mu}^a T_a + i g_x Q_x B_{\mu}^x, \qquad g_x = \frac{g^{\prime}}{\sqrt{1-t_w^2/3}}
\end{equation}
with $t_{W} \equiv \tan{\te_{W}}$, and $g$, $g^{\prime}$ the SM $SU(2)_{L} \otimes U(1)_{Y}$ gauge couplings.

The global symmetries prevent the appearance of a Higgs potential at
tree level. The Higgs potential is indeed generated at one-loop and
higher orders due to the interactions with gauge bosons and fermions
through the CW potential. One can show \cite{Cheung:2006nk} that in
this setup the pseudo-scalar $\eta$ remains massless, while the Higgs
boson acquires a mass through one-loop logarithmic and two-loop
quadratic divergences (this is due to the collective symmetry breaking
mechanism). To force $\eta$ to have a non-zero mass, in order to avoid
a new and not observed long-range interaction, one possible solution
is to introduce a term 
\begin{equation}
	-\mu_{\phi}^{2} \left( \phi_{1}^{\dagger} \phi_{2} \text{ +h.c.} \right)
\end{equation}
into the CW potential by hand. This explicitly breaks the global
$SU(3)$ symmetry and also the collective symmetry breaking mechanism,
but the corrections are small \cite{Cheung:2006nk}: we will adopt this
extension, and the parameter $\mu_{\phi}$ will be then proportional to
the pseudo-scalar mass $m_{\eta}$. The CW potential now becomes 
\begin{equation}
	V_{CW} = - \mu^{2} h^{\dagger} h + \lambda ( h^{\dagger} h
        )^{2} - \frac{1}{2} m_{\eta}^{2} \eta^{2}+ \lambda^{\prime}
        h^{\dagger} h \eta^{2} + \ldots 
	\label{SLHCW}
\end{equation}
where the parameters are defined as in
\cite{Cheung:2006nk}. From the minimization of the potential one
obtains the expression for the \emph{vev} of the Higgs field  
\begin{equation}
	v^{2} = \frac{\mu^{2}}{\lambda}
	\label{veq}
\end{equation}
and the mass of the pseudo-scalar
$\eta$
\begin{equation}
	m_{\eta}^{2} = \frac{\mu_{\phi}^{2}}{c_{\be} s_{\be}} \cos{\left(\frac{v}{\sqrt{2} f s_{\be} c_{\be}} \right)}.
	\label{masseta}
\end{equation}

If one assumes that the physics at the cut-off $\Lambda = 4 \pi f$
gives no sizable contribution to the scalar potential, then the CW
potential ($\ref{SLHCW}$) fully determines the scalar masses and their
couplings: in order to realize a correct EWSB pattern, the free
parameters of the CW potential are then not anymore independent among
themselves. In particular, we require the parameter $\mu$ to reproduce
the observed Higgs boson mass 
\begin{equation}
	m_h = \sqrt{2} \, \mu
	\label{Higgsmass}
\end{equation}
while we fix \emph{v} in Eq. ($\ref{veq}$) in order to match the
prediction of the SM $W$-boson mass: the $W$-boson mass is indeed
predicted to be \cite{Cheung:2006nk} 
\begin{equation}
	m_{W} = \frac{g v}{2} \left[ 1-\frac{1}{12}
          \frac{v^{2}}{f^{2}} \frac{\tb^{4}-\tb^{2}+1}{\tb^{2}} +
          \frac{1}{180} \frac{v^{4}}{f^{4}}
          \frac{\tb^{8}-\tb^{6}+\tb^{4}-\tb^{2}+1}{\tb^{4}} +
          \mathcal{O} \left( \frac{v^{6}}{f^{6}} \right) \right] 
\end{equation}
and therefore we require $v$ to satisfy 
\begin{eqnarray}
	v &\simeq& v_{SM} \left[ 1+\frac{1}{12}
          \frac{v^{2}_{SM}}{f^{2}} \frac{\tb^{4}-\tb^{2}+1}{\tb^{2}} -
          \frac{1}{180} \frac{v^{4}_{SM}}{f^{4}}
          \frac{\tb^{8}-\tb^{6}+\tb^{4}-\tb^{2}+1}{\tb^{4}}
        \right] \label{vdef} \\ 
	&\equiv& v_{SM} \left[ 1+ \delta_{v}^{(2)}-\delta_{v}^{(4)} \right] \nonumber
\end{eqnarray}
where $v_{SM} = 246 \text{ GeV}$, in order to have $m_{W}= g \,
v_{SM}/2$. Therefore ($\ref{Higgsmass}$) and ($\ref{veq}$,
$\ref{vdef}$) are two conditions which have to be imposed on the free
parameters of the CW potential, i.e. on $f$, $\tb$, $\mu_{\phi}$, $R$
($R$ is a ratio of Yukawa couplings of the fermion sector which
affects $\mu$, $\lambda$, cf. below): we decided to let $f$ and $\tb$
to be free parameters of our study, fixing the values of $\mu_{\phi}$
and $R$ through the previous equations. 

From ($\ref{vdef}$) we also see that the correction to $v_{SM}$ is
proportional to $\tb^{2} v_{SM}^{2}/f^{2}$ in the large $\tb$ limit:
as suggested in \cite{Cheung:2006nk}, for perturbation theory to be valid,
the $\mathcal{O} \left( v_{SM}^{4}/f^{4} \right)$ correction should be
suppressed by a factor of 0.1 relative to the $\mathcal{O} \left(
  v_{SM}^{2}/f^{2} \right)$ correction, i.e.  
\begin{equation}
	\delta_{v}^{(4)}/\delta_{v}^{(2)}<0.1.
	\label{perturbative}
\end{equation} 
We will require this latter condition to be satisfied in the
considered parameter space of the model. 

After EWSB and using relation ($\ref{vdef}$), the leading order mass
spectrum of the heavy and light (SM) gauge bosons is given by
\cite{Han:2005ru,Cheung:2006nk} 
\begin{eqnarray}
	m_{W^{\pm}} = \frac{g \, v_{SM}}{2} & \qquad & m_{X^{\pm}} =
        m_{Y^{0}} = m_{\bar{Y}^{0}} = \frac{g f}{\sqrt{2}} \nonumber
        \\ 
	m_{Z} = \frac{g \, v_{SM}}{2 c_w} \left( 1+ \frac{v^2}{16 f^2}
          (1-t_w^2)^2 \right) & \qquad & m_{Z^{\prime}} =
        \sqrt{\frac{2}{3-t_{w}^{2}}} g f \\ 
	m_{\gamma}=0 \nonumber
\end{eqnarray}
where we have included also the $\mathcal{O}(v^2/f^2)$ custodial
symmetry violating shift term in the $Z$-mass, and $c_w$ is the cosine
of the Weinberg angle. If we parametrize the interaction terms of the
charged gauge bosons to the Higgs field in the following way 
\begin{equation}
	\mathcal{L}_{\Phi} \supset 2 \, \frac{m_{W}^{2}}{v} \, y_{W}
        \, W^{+} W^{-} h+2 \, \frac{m_{X}^{2}}{v} \, y_{X} \, X^{+}
        X^{-} h 
\end{equation}
then the couplings $y_{V}$ after EWSB are predicted to be
\cite{Wang:2011rv} 
\begin{equation}
	y_{W} \simeq \frac{v}{v_{SM}} \left[ 1- \frac{1}{4}
          \frac{v_{SM}^{2}}{f^{2}} \frac{\tb^{4}-\tb^{2}+1}{\tb^{2}} +
          \frac{1}{36} \frac{v_{SM}^{4}}{f^{4}} \frac{\left( \tb^{2}-1
            \right)^{2}}{\tb^{2}} \right], \qquad y_{X} \simeq
        -\frac{1}{2} \frac{v^{2}}{f^{2}} \; .
\end{equation}
Using relation ($\ref{vdef}$), we can express the corrections of the
$hVV$ ($V \equiv W,Z$) couplings with respect to their SM value up to
$\mathcal{O} \left( v_{SM}^4/f^4 \right)$ equivalently as
\cite{Cheung:2006nk} 
\begin{eqnarray}
	\frac{g_{hWW}}{g_{hWW}^{SM}} &=& 1- \frac{1}{4}
        \frac{v_{SM}^{2}}{f^{2}} \left(
          \frac{\tb^{4}-\tb^{2}+1}{\tb^{2}} \right) + \frac{1}{36}
        \frac{v_{SM}^{4}}{f^{4}} \frac{\left( \tb^{2}-1
          \right)^{2}}{\tb^{2}} \label{ghVVSLH} \\ 
	\frac{g_{hZZ}}{g_{hZZ}^{SM}} &=& 1- \frac{1}{4}
        \frac{v_{SM}^{2}}{f^{2}} \left(
          \frac{\tb^{4}-\tb^{2}+1}{\tb^{2}} + \left( 1-t_{w}^{2}
          \right)^{2} \right) + \frac{1}{36} \frac{v_{SM}^{4}}{f^{4}}
        \frac{\left( \tb^{2}-1 \right)^{2}}{\tb^{2}} \nonumber 
\end{eqnarray}
where as usual
\begin{equation}
	g_{hVV}= \frac{m_{V}^{2}}{v} \, y_{V}, \qquad g_{hVV}^{SM} = \frac{m_{V}^2}{v}_{\big{|_{v=v_{SM}}}}.
\end{equation}


\subsection{Fermion sector}

\subsubsection*{Littlest Higgs}

The SM fermions acquire their masses through the Higgs mechanism via
Yukawa interactions: the large top Yukawa coupling induces a dominant
quadratic correction to the Higgs boson mass, spoiling the naturalness
of a light Higgs boson. In $L^2H$ model this problem is solved by
introducing a new set of heavy fermions with coupling to the Higgs
field such that it cancels the quadratic divergence due to the top
quark. The new fermions are a vectorlike pair $\left( T^{\prime},
  T^{\prime \, c} \right)$ with quantum numbers
$(\textbf{3},\textbf{1})_{Y_i}$,
$(\bar{\textbf{3}},\textbf{1})_{-Y_i}$ respectively, and therefore
they are allowed to have a bare mass term which is chosen to be of
order $f$.  

The Yukawa-like Lagrangian for the third generation of quarks can be
found e.g. in~\cite{Han:2003wu}, and contains the following interaction
terms to the Higgs after EWSB: 
\begin{equation}
	\mathcal{L}_{t} \supset -\lambda_{1} f \left(
          \frac{s_{\Sigma}}{\sqrt{2}} \, \bar{t}^{\prime}_L \,
          t_{R}^{\prime} + \frac{1+c_{\Sigma}}{2} \,
          \bar{T}^{\prime}_L \, t_{R}^{\prime} \right)- \lambda_{2} f
        \bar{T}^{\prime}_{L} \, T^{\prime}_{R} \text{ +h.c.} 
	\label{topLH}
\end{equation}
where $c_{\Sigma}= \cos{\left(\sqrt{2} h/f \right)}$, $s_{\Sigma}=
\sin{\left(\sqrt{2} h/f \right)}$, and with $\lambda_{1,2}$ as free
parameters. After diagonalization of the mass matrix, the leading
order mass eigenvalues are the following 
\begin{equation}
	m_t = \frac{\lambda_2 R}{\sqrt{1+R^2}} v, \qquad m_T =
        \lambda_2 \sqrt{1+R^2} f \quad. 
\end{equation}
Here, we have defined the ratio of the Yukawa couplings
\begin{equation}
	R= \lambda_1/\lambda_2.
	\label{RdefLH}
\end{equation}
However we can fix $\lambda_{2}$ requiring that, for given $\left( f,R
\right)$, $m_{t}$ corresponds to the experimental top mass value: in
this way, the only free parameters in the top sector are $f$ and
$R$. If we parametrize the interaction terms of the top quark and
heavy top to the Higgs field (the dominant contributions to the effective
$hgg$ vertex) in the following way 
\begin{equation}
	\mathcal{L}_{t} \supset -\frac{m_{t}}{v} \, y_{t} \, \bar{t}
        \, t \, h- \frac{m_{T}}{v} \, y_{T} \, \bar{T} \, T \, h
        \qquad ,
\end{equation}
then the couplings $y_{t,T}$ after EWSB are predicted up to $\mathcal{O} \left(v^{2}/f^{2} \right)$ to be \cite{Wang:2011rv}
\begin{equation}
	y_{t} = 1 - \frac{v^2}{f^2} \left[ \frac{2 R^4+R^2+2}{3(1+R^2)^2} + \frac{x^2}{4}-\frac{x}{2} \right], \qquad y_{T} = -\frac{R^2}{(1+R^2)^2} \frac{v^2}{f^2}.
\end{equation}

The scalar interactions with the \emph{up}-type quarks of the first
two generations have the same form as $\mathcal{L}_t$, except that
there is no need for extra vectorlike quarks. The interactions with
the \emph{down}-type quarks and leptons of the three generations are
generated by a similar Lagrangian, again without the extra vectorlike
quarks. For the explicit forms of the Lagrangian terms we refer as
before to Ref. \cite{Han:2003wu}. The important result for our
analysis is the explicit correction of the Higgs-fermion couplings
with respect to their SM value: from the Feynman rules of the vertices
$huu$ and $hdd$ listed in the appendix of Ref. \cite{Han:2003wu}, and
using relation ($\ref{vevLH}$), we obtain up to $\mathcal{O} \left(
  v_{SM}^2/f^2 \right)$ 
\begin{equation}
	\frac{g_{hff}}{g_{hff}^{SM}} = 1 - \frac{1}{2}\frac{v_{SM}^2}{f^2} \left[ \frac{7}{4} +\frac{x^2}{4} -x \right] \qquad f \equiv u,d,c,s,b
	\label{hffLH}
\end{equation}
where 
\begin{equation}
	g_{hff}= \frac{m_{f}}{v} \, y_{f}, \qquad g_{hff}^{SM} = \frac{m_{f}}{v}_{\big{|_{v=v_{SM}}}}.
\end{equation}
Eq. ($\ref{hffLH}$) will allow us to calculate the tree-level decays
of the Higgs boson into two fermions, cf. Appendix~\ref{app:higgs}.

\subsubsection*{Littlest Higgs with \emph{T}-parity}

To implement \emph{T}-parity in the fermion sector one introduces two
$SU(2)_{A}$ fermion doublets $q_{A} =\left( i d_{L_{A}},-i u_{L_{A}}
\right)^{T}$ with $A=1,2$, as in \cite{Chen:2006cs}: \emph{T}-parity
will be defined such that $q_{1} \leftrightarrow -q_{2}$. The
\emph{T}-even combination $u_{L+}=\left( u_{L_1}-u_{L_2}
\right)/\sqrt{2}$ will be the \emph{up}-type component of the SM
fermion doublet, while the \emph{T}-odd combination $u_{L-} = \left(
  u_{1}+u_{2} \right)/\sqrt{2}$ will be its \emph{T}-odd partner: the
same definitions hold also for the \emph{down}-type components. 

We require that the \emph{T}-even (SM) eigenstates obtain a mass only
from Yukawa-like interactions after EWSB, while forcing the masses of
the \emph{T}-odd eigenstates to be at the TeV scale. A possible
Lagrangian that could generate a TeV mass only for the \emph{T}-odd
combinations can be found in~\cite{Chen:2006cs}: 
\begin{eqnarray}
	\mathcal{L}_{k} \supset &-& \sqrt{2} k f \left[ \bar{d}_{L-}
          \, \tilde{d_{c}} + \frac{1+c_{\xi}}{2} \, \bar{u}_{L-} \,
          \tilde{u}_{c} - \frac{s_{\xi}}{\sqrt{2}} \, \bar{u}_{L-} \,
          \chi_{c} - \frac{1-c_{\xi}}{2} \, \bar{u}_{L-} \, u_{c}
        \right]+ \nonumber \\  
	&-& m_{q} \, \bar{u}^{\prime}_{c} \, u_{c}- m_{q} \,
        \bar{d}^{\prime}_{c} \, d_{c} - m_{\chi} \,
        \bar{\chi}^{\prime}_{c} \, \chi_{c} \quad \text{+ h.c.} 
	\label{Todd}
\end{eqnarray}
where $c_{\xi}= \cos{\left( h / \sqrt{2} f \right)}$, $s_{\xi}=
\sin{\left( h / \sqrt{2} f \right)}$. $u_{L-}$ and $d_{L-}$ are the
\emph{T}-odd eigenstates, while the other fields $u_c$, $d_c$,
$\tilde{u}_c$, $\tilde{d}_c$, $u^{\prime}_c$, $d^{\prime}_c$,
$\chi_c$, $\chi^{\prime}_c$ are all embedded in the so called
\emph{mirror fermions} necessary to write down an invariant Lagrangian
under all symmetries. $k$, $m_q$ and $m_{\chi}$ are matrices in
flavor space for both quarks and leptons: we will assume for
simplicity that these matrices are all diagonal and flavor
independent. 

One can notice that the \emph{down}-type fermions have only Dirac-mass terms and no interactions with the Higgs: 
\begin{equation}
	-\sqrt{2} k f \, \bar{d}_{L-} \, \tilde{d}_{c}-m_{q} \, \bar{d}_{c}^{\prime} \, d_{c}.
\end{equation}
They are thus already mass eigenstates with masses
\begin{equation}
	m_{1} = \sqrt{2} k f, \qquad	m_{2} = m_{q}		
\end{equation}
and their contributions will not be considered in the effective one-loop couplings of the Higgs, since they do not couple to the Higgs at tree level.

On the other side, the \emph{up}-type combinations in ($\ref{Todd}$)
have Dirac-mass terms and also couplings with the Higgs ($c_{\xi}$ and
$s_{\xi}$): by diagonalizing these couplings, one obtains the following
mass spectrum at leading order 
\begin{equation}
	m^{h}_{1} = \sqrt{2} k f, \qquad m^{h}_{2} = m_{\chi}, \qquad m^{h}_{3} = m_{q}		
\end{equation}
where the superscript $h$ indicates that the eigenstates also have an
interaction with the Higgs field. The resulting couplings with the
Higgs up to $\mathcal{O} \left(v^2/f^{2} \right)$ are 
\begin{equation}
	\mathcal{L}_{k} \supset - \frac{m_{1}^{h}}{v} \, y_{1} \,
        \bar{u}_{1} \, u_{1} \, h- \frac{m_{2}^{h}}{v} \, y_{2} \,
        \bar{u}_{2} \, u_{2} \, h- \frac{m_{3}^{h}}{v} \, y_{3} \,
        \bar{u}_{3} \, u_{3} \, h 
\end{equation}
with
\begin{equation}
	y_{1} = -\frac{1}{4} \frac{v^{2}}{f^{2}} \frac{1}{1- \frac{2
            \, f^{2} \, k^{2} }{m_{\chi}^{2}}}, \qquad y_{2} = -\frac{
          k^{2} \, v^{2} }{ 2 \, m_{\chi}^{2} } \frac{1}{1- \frac{2 \,
            f^{2} \, k^{2} }{m_{\chi}^{2}}}, \qquad y_{3} =
        \mathcal{O} \left( \frac{v^{4}}{f^{4}} \right). 
	\label{tildetyuk}
\end{equation}

We can further reduce the number of free parameters assuming that $m_{q}$ and $m_{\chi}$ are large enough such that the Higgs couplings ($\ref{tildetyuk}$) are independent from their values up to $\mathcal{O} \left( v^{2}/f^{2} \right)$, i.e.
\begin{table}[!ht]
\centering
\begin{tabular}{c|c||c}
mass (\emph{up}-type) & Higgs coupling (only \emph{up}-type) & mass (\emph{down}-type) \\ \hline & & \\
$m^{h}_{1,i} = \sqrt{2} k f \quad$ & $y^{i}_{1} = -\frac{1}{4} \frac{v^{2}}{f^{2}} \quad$ & $m_{1,i} = \sqrt{2} k f$ \\ & & \\
$m^{h}_{2,i} = m_{\chi} \quad$ & $y^{i}_{2} = 0 \quad$ & $m_{2,i} = m_{q}$ \\ & & \\
$m^{h}_{3,i} = m_{q} \quad$ & $y^{i}_{3} = \mathcal{O} \left( v^{4}/f^{4} \right)$ & 
\end{tabular}
\end{table}%
\\where we have restored the flavor index $i=1,2,3$ referring to both quarks and leptons.

Under these assumptions, in the effective one-loop couplings of the
Higgs we will then consider only the contributions from the three
degenerate \emph{up}-type \emph{T}-odd quarks $u_{1}^i$, since the
other couplings are either suppressed ($y_{2}^i$, $y_{3}^i$) or absent
(\emph{down}-type). The \emph{T}-odd heavy neutrinos coming from the
same interactions are clearly not included in the couplings of the
Higgs with gluons and photons, being neither colored, nor electrically
charged. 

These new twelve \emph{T}-odd partners $u_1^i$, $d_1^i$ of the SM
fermions can also generate four-fermion operators via box diagrams
involving the exchange of NGBs \cite{Hubisz:2005tx}. Assuming always
that the couplings $k$ are flavor-diagonal and flavor-independent,
the generated operators have the form 
\begin{equation}
	\mathcal{O}_{\text{4-f}} = - \frac{k^2}{128 \, \pi^2 f^2}
        \bar{\psi}_L \gamma^{\mu} \psi_L \bar{\psi}^{\prime}_L
        \gamma_{\mu} \psi^{\prime}_L + \mathcal{O} \left( \frac{g}{k}
        \right), 
\end{equation}
where $\psi$ and $\psi^{\prime}$ are (distinct) SM fermions. The
experimental bound on four-fermion interactions involving SM fields
provides an upper bound on the \emph{T}-odd fermion masses: the
strongest constraint comes from the \emph{eedd} operator, whose
coefficient is required to be smaller than $2 \pi /(26.4 \text{
  TeV})^2$~\cite{Hubisz:2005tx,Beringer:1900zz}, which thus yields 
\begin{equation}
	k^2 \lesssim 0.367 \, \pi^3 \, f_{\text{TeV}}^2
	\label{kbound}
\end{equation}
where $f_{\text{TeV}}$ is the value of $f$ in units of TeV. Taking a
closer look to the contribution of the \emph{T}-odd fermions $u_1^i$
to the signal strength modifier, one can notice that their
contribution enters only in the combination  
\begin{equation}
	F_{1/2} (m_{1}^{h}) \cdot y_{1}
\end{equation}
in the expression of the partial decay widths of the Higgs into two
gluons and photons, cf. Eq. ($\ref{hgg}$) and ($\ref{hgaga}$),
respectively. However the coupling $y_{1}$ is independent of $k$ at
the order we are considering, and the loop factor $F_{1/2}$ approaches
a constant value $F_{1/2} \rightarrow -4/3$ when the particle in the
loop is much heavier than the Higgs \cite{Gunion:1989we}, as in our
case ($m_{1}^{h} \gg m_h$): the net contribution of the heavy
\emph{T}-odd fermions is thus in good approximation independent of the
value of $k$. Without loss of generality we could therefore choose $k$
to saturate the four-fermion interaction bound ($\ref{kbound}$), with
an upper limit of 4$\pi$ when $f \rightarrow \infty$. 

The next task is to write invariant Yukawa-like terms to give mass to
the \emph{T}-even (SM) combinations $u_{L+}$ and $d_{L+}$. In order to
avoid dangerous contributions to the Higgs mass from one-loop
quadratic divergences, the top Yukawa sector must also incorporate a
collective symmetry breaking pattern.  

The details of the procedure could again be found in
\cite{Chen:2006cs}: the Yukawa-like Lagrangian for the top sector
contains the following terms 
\begin{equation}
	\mathcal{L}_{t} \supset -\lambda_{1} f \left(
          \frac{s_{\Sigma}}{\sqrt{2}} \, \bar{t}_{L+} \,
          t_{R}^{\prime} + \frac{1+c_{\Sigma}}{2} \,
          \bar{T}^{\prime}_{L+} \, t_{R}^{\prime} \right)- \lambda_{2}
        f \left( \bar{T}^{\prime}_{L+} \, T^{\prime}_{R+} +
          \bar{T}^{\prime}_{L-} \, T^{\prime}_{R-} \right) \text{
          +h.c.} 
	\label{topLHT}
\end{equation}
where $c_{\Sigma}= \cos{\left( \sqrt{2} h/f \right)}$ and $s_{\Sigma}=
\sin{\left( \sqrt{2} h/f \right)}$. 

Among the terms that we have neglected, there are the interaction
terms of the \emph{T}-odd eigenstate $t_{L-}$, which does not acquire
any mass term from $\mathcal{L}_{t}$ while obtaining its mass from
$\mathcal{L}_{k}$ as explained before. In $\mathcal{L}_t$ a different
\emph{T}-odd Dirac fermion $T_{-} \equiv \left(
  T^{\prime}_{L-},T^{\prime}_{R-} \right)$ obtains a high-scale mass 
\begin{equation}
	m_{T_{-}} = \lambda_{2} \, f \ .
\end{equation}
It does not have tree-level interactions to the Higgs boson, and will
be thus not included in the Higgs one-loop effective couplings.   

One should notice that the \emph{T}-even top Lagrangian
($\ref{topLHT}$) has the same form as the $L^2H$ top Lagrangian
($\ref{topLH}$): the mass spectrum and the couplings to the Higgs
boson will therefore be the same in both models, by simply setting
$x=0$. The \emph{T}-even combinations in $\mathcal{L}_{t}$,
i.e. $\left( t_{L+}, t_{R}^{\prime} \right)$ and $\left(
  T_{L+}^{\prime}, T_{R+}^{\prime} \right)$, mix among each other: 
\begin{equation}
	-\mathcal{L}_t \supset \left( \begin{array}{cc} \bar{t}_{L+} &
            \bar{T}_{L+}^{\prime} \end{array} \right) \, \mathcal{M}
        \left( \begin{array}{c} t_R^{\prime} \\
            T_{R+}^{\prime} \end{array} \right) \text{ +h.c.,} \qquad
        \mathcal{M} = \left( \begin{array}{cc} \frac{ \lambda_{1}
              f}{\sqrt{2}} \sin{\frac{\sqrt{2} h}{f}} & 0 \\ & \\
            \lambda_{1} f \cos^2{\frac{h}{\sqrt{2} f}} & \lambda_2
            f \end{array} \right). 
	\label{matrix}
\end{equation}
The mass terms are diagonalized by defining the linear combination
\cite{Hubisz:2005tx} 
\begin{eqnarray}
	t_L &=& \cos{\be} \cdot t_{L+} - \sin{\be} \cdot
        T^{\prime}_{L+}, \qquad T_{L+} = \sin{\be} \cdot t_{L+}
        +\cos{\be} \cdot T^{\prime}_{L+} \nonumber \\ t_R &=&
        \cos{\al} \cdot t_R^{\prime} - \sin{\al} \cdot T^{\prime}_{L+}
        \qquad T_{R+} = \sin{\al} \cdot t_R^{\prime} + \cos{\al} \cdot
        T_{R+}^{\prime} 
\end{eqnarray}
Here, we use the dimensionless ratio $R=\lambda_1/\lambda_2$ as well
as the leading order expressions of the mixing angles 
\begin{equation}
	\sin \alpha = \frac{R}{\sqrt{1+R^2}}, \qquad \sin \be =
        \frac{R^2}{1+R^2} \frac{v}{f}. 
\end{equation}
The leading order mass spectrum is the following
\begin{equation}
	m_{t} = \frac{\lambda_{2} R}{\sqrt{1+R^{2}}} v, \qquad
        m_{T_+} = \lambda_{2} \sqrt{1+R^{2}} f \quad .
\end{equation}
Again, $R$ and $\lambda_{2}$ are considered to be free
parameters. However we can fix $\lambda_{2}$ requiring that, for given
$\left( f,R \right)$, $m_{t}$ corresponds to the experimental top mass
value: this way, the only free parameters in the \emph{T}-even top
sector are $f$ and $R$.  

The resulting couplings to the Higgs up to $\mathcal{O} \left(
  v^{2}/f^{2} \right)$ are given by \cite{Belyaev:2006jh,Wang:2011rv} 
\begin{equation}
	\mathcal{L}_{t} \supset -\frac{m_{t}}{v} \, y_{t} \, \bar{t}
        \, t \, h- \frac{m_{T_{+}}}{v} \, y_{T_{+}} \, \bar{T}_{+} \,
        T_{+} \, h 
\end{equation}
with 
\begin{equation}
	y_{t}=1-\frac{v^{2}}{f^{2}} \frac{2 R^{4}+R^{2}+2}{3\left(
            1+R^{2} \right)^{2}}, \qquad y_{T_{+}} = -
        \frac{v^{2}}{f^{2}} \frac{R^{2}}{\left( 1+R^{2} \right)^{2}}. 
	\label{topyuk}
\end{equation}

The other two generations of \emph{T}-even (SM-like) \emph{up}-type
quarks acquire their mass through analogous terms as
$\mathcal{L}_{t}$, but with the $T_{\pm}$ missing since the Yukawa
couplings are small and one does not have to worry about the quadratic
divergences. Using Eq. ($\ref{vevLHT}$), the corrections to the Yukawa
couplings with respect to their SM values up to $\mathcal{O} \left(
  v_{SM}^4/f^4 \right)$ are given by \cite{Chen:2006cs} 
\begin{equation}
	\frac{g_{h \bar{u} u}}{g_{h \bar{u} u}^{SM}} = 1-\frac{3}{4}
        \frac{v_{SM}^{2}}{f^{2}}-\frac{5}{32} \frac{v_{SM}^{4}}{f^{4}}
        \qquad u \equiv u,c. 
\end{equation}

We need to construct also a Yukawa interaction which gives a mass
after EWSB to the \emph{T}-even (SM-like) \emph{down}-type quarks and
charged leptons. Two possible constructions of Lagrangians can be
found in~\cite{Chen:2006cs}, which will be denoted as \emph{Case A}
and \emph{Case B}, respectively. The corresponding corrections to the
Yukawa couplings with respect to their SM values up to $\mathcal{O}
\left( v_{SM}^4/f^4 \right)$ are given by ($d \equiv d,s,b,l^{\pm}_i$) 
\begin{eqnarray}
	\frac{g_{h \bar{d} d}}{g_{h \bar{d} d}^{SM}} &=& 1-
        \frac{1}{4} \frac{v_{SM}^{2}}{f^{2}} + \frac{7}{32}
        \frac{v_{SM}^{4}}{f^{4}} \qquad \text{Case A} \nonumber \\ 
	\frac{g_{h \bar{d} d}}{g_{h \bar{d} d}^{SM}} &=& 1-
        \frac{5}{4} \frac{v_{SM}^{2}}{f^{2}} - \frac{17}{32}
        \frac{v_{SM}^{4}}{f^{4}} \qquad \text{Case B}. 
	\label{dcoupling}
\end{eqnarray}
We will analyze the parameter space of the \emph{LHT} model with both
\emph{Case A,B} implementations: it is to be noted that \emph{Case B}
predicts a stronger suppression for the down-type fermion couplings to
the Higgs boson, and this will have an influence on our results.

\subsubsection*{Simplest Little Higgs}

Since this model contains a gauged $SU(3)$, SM fermions that are
doublets under $SU(2)$ must be enlarged into triplets under
$SU(3)$. In addition, new $SU(3)$ singlet fermions must be introduced
to cancel the hypercharge anomalies and to give mass to the new third
components of the $SU(3)$ triplet fermions. In the ``anomaly-free''
scenario, the quarks of the third generation and all leptons are
embedded into \textbf{3} of $SU(3)$: 
\begin{equation}
	Q_{3}^{T} = \left( t,b,iT \right), \qquad L_{m}^{T} = \left(
          \nu_{m}, l_{m}, i N_{m} \right) \quad (m=1,2,3) 
\end{equation}
adding also the corresponding right handed singlets $i t^{c}$,
$ib^{c}$, $i T^{c}$ and $i l_{m}^{c}$, $i N_{m}^{c}$. We do not
include a right-handed neutrino, leaving the neutrinos as massless.  

The corresponding Yukawa Lagrangian $\mathcal{L}_{Y}$ can be found
explicitly in~\cite{Han:2005ru,delAguila:2011wk}, and gives rise at
leading order to the following mass spectrum after EWSB: 
\begin{table}[!ht]
\centering
\begin{tabular}{ccc}
$m_{b} \propto \lambda^b \quad$ & $m_{t} = \lambda_{2}^{t} \, v \, R \sqrt{\frac{\tb^{2}+1}{2(\tb^{2}+R^{2})}} \quad$ & $m_{T} = \lambda_{2}^{t} \, f \sqrt{\frac{\tb^{2}+R^{2}}{\tb^{2}+1}}$ \\ \\
$m_{\nu} = 0 \quad$ & $m_{l_{m}} \propto \lambda^{l}_{m,n} \quad$ & $m_{N_{m}} = \lambda_{N_{m}} \, f \frac{\tb}{\sqrt{1+\tb^{2}}}$ 
\end{tabular}
\end{table}%
The free parameters are $R =
\left(\lambda^{t}_{1}/\lambda_{2}^{t}\right)$, $\lambda_{2}^{t}$,
$\lambda^{b}$ $f$, $\tb$ in the 
top sector, and $\lambda^{N_{m}}$, $\lambda^{l}_{m,n}$, $f$, $\tb$ in
the lepton sector, respectively. However we can fix $\lambda_{2}^{t}$, $\lambda^{b}$
and $\lambda^{l}_{m,n}$ requiring that for given ($f,\tb$) the
predicted values of $m_{t}$, $m_{b}$ and $m_{l_{m}}$ correspond to
their experimental values: in this way the free parameters in the top
sector are $f$ and $\tb$ (notice that $R$ is fixed by the EWSB
requirement), while in the lepton sector they are $\lambda_{N_{m}}$, $f$,
$\tb$. Since the heavy neutrinos do not affect neither the effective
$hgg$ coupling, nor the $h\gamma \gamma$ one, the parameter
$\lambda_{N_{m}}$ is thus irrelevant for our study. 

Regarding the couplings of these fermions to the Higgs, they can be
parametrized as 
\begin{equation}
	\mathcal{L}_{Y} \supset - \frac{m_{l}^{i}}{v} \, y_{l}^{i} \,
        \bar{l}_{i} \, l_{i} \, h - \frac{m_{N}^{i}}{v} \, y_{N}^{i}
        \, \bar{N}_{i} \, N_{i} \, h - \frac{m_{t}}{v} \, y_{t} \,
        \bar{t} \, t \, h - \frac{m_{T}}{v} \, y_{T} \, \bar{T} \, T
        \, h -\frac{m_{b}}{v} \, y_{b} \, \bar{b} \, b \, h 
\end{equation}
and predicted up to $\mathcal{O} \left( v^2/f^2 \right)$ to
be~\cite{Han:2005ru}  
\begin{eqnarray}
	y_{l}^{i} &=& 1- \frac{1}{6} \frac{v^{2}}{f^{2}} \left( 3+
          \frac{\tb^{4}-\tb^{2}+1}{\tb^{2}} \right), \qquad y_{N}^{i}
        = \mathcal{O} \left( \frac{v^{4}}{f^{4}} \right) \nonumber \\ 
	y_{t} &=& 1- \frac{1}{6} \frac{v^{2}}{f^{2}} \left[
          \frac{\left( 1+\tb^{2} \right)^{2} \left( R^{4}-R^{2} \,
              \tb^{2} + \tb^{4} \right)}{\tb^{2} \left( \tb^{2} +
              R^{2} \right)^{2}} \right], \qquad y_{T} = - \frac{1}{2}
        \frac{v^{2}}{f^{2}} R^{2} \left(
          \frac{\tb^{2}+1}{\tb^{2}+R^{2}} \right)^{2} \nonumber \\ 
	y_{b} &=& 1- \frac{1}{6} \frac{v^{2}}{f^{2}} \left( 3+
          \frac{\tb^{4}-\tb^{2}+1}{\tb^{2}}
        \right). \label{tSLHcoupling} 
\end{eqnarray}
The corrections of the bottom-quark and lepton Yukawa couplings
with respect to their SM values up to $\mathcal{O} \left( v_{SM}^4/f^4 \right)$
are equivalently given by 
\begin{equation}
	\frac{g_{h\bar{b}b}}{g_{h\bar{b}b}^{SM}} = \frac{g_{h
            \bar{l}l}}{g_{h\bar{l}l}^{SM}} = 1- \frac{1}{4}
        \frac{v_{SM}^{2}}{f^{2}} \left(
          \frac{\tb^{4}+\tb^{2}+1}{\tb^{2}} \right) - \frac{1}{720}
        \frac{v_{SM}^{4}}{f^{4}} \left( \frac{\tb^{8}+24 \tb^{6}-19
            \tb^{4}+24 \tb^{2}+1}{\tb^{4}} \right) 
\end{equation}
where as usual 
\begin{equation}
	g_{hff}= \frac{m_{f}}{v} \, y_{f}, \qquad g_{hff}^{SM} =
        \frac{m_{f}}{v}_{\big{|_{v=v_{SM}}}}. 
\end{equation}

In the ``anomaly free'' embedding, the first two generations of quarks
are embedded into $\textbf{3}^{*}$ of $SU(3)$ with the corresponding
right-handed singlets: 
\begin{eqnarray*}
	Q_{1}^{T} &=& \left( d,-u,i D \right) \qquad id^{c}, i u^{c}, i D^{c} \\
	Q_{2}^{T} &=& \left( s, -c, i S \right) \qquad i s^{c}, i c^{c}, i S^{c}
\end{eqnarray*}
Notice that the heavy vector-like quarks of the first two generations
have electric charge $-1/3$ in contrast to the charge $+2/3$ of the
heavy quark of the third generation. The Lagrangian terms for the
Yukawa couplings of the first two generations of quarks can be found
in \cite{Han:2005ru}, and the resulting mass spectrum after EWSB
consists of the SM quarks plus two heavy $D$, $S$ quarks with charge
$-1/3$.  

As suggested in~\cite{Han:2005ru}, one would expect an $h \, D^{c}_{m}
\, D_{m}$ coupling at order $v/f$ as for the top sector, but this term
is exactly canceled by the contribution from $h \, D^{c}_{m} \, d_{m}$
after $d$-$D$ mixing if the down and strange quark masses are
neglected. For this reason we will not include the heavy $D$, $S$ in
the calculation of the one-loop effective couplings: therefore only
the contributions from the top (the dominant one among the SM-like
particles) and from the heavy top $T$ will be included in the one-loop
effective couplings. 

The corrections of the charm and strange quarks Yukawa couplings
with respect to their SM values up to $\mathcal{O} \left( v_{SM}^4/f^4 \right)$
are finally given by 
\begin{equation}
	\frac{g_{h\bar{c}c}}{g_{h\bar{c}c}^{SM}} = \frac{g_{h
            \bar{s}s}}{g_{h\bar{s}s}^{SM}} = 1- \frac{1}{4}
        \frac{v_{SM}^{2}}{f^{2}} \left(
          \frac{\tb^{4}+\tb^{2}+1}{\tb^{2}} \right) - \frac{1}{720}
        \frac{v_{SM}^{4}}{f^{4}} \left( \frac{\tb^{8}+24 \tb^{6}-19
            \tb^{4}+24 \tb^{2}+1}{\tb^{4}} \right). 
\end{equation}


\subsection{Electroweak Precision Observables}

\subsubsection*{Littlest Higgs}

The contribution to EWPO from the $L^2H$ structure was already studied
in Ref. \cite{Csaki:2002qg}. They calculated all contributions to EWPO
from the tree-level exchange of the heavy gauge bosons and from the
presence of the triplet \emph{vev} $v^{\prime}$: there should in
principle be also contributions due to heavy quark loop modifications
to the light gauge boson propagators, but as the authors of
Ref.~\cite{Csaki:2002qg} have shown, these contributions are almost an
order of magnitude smaller than the maximal contribution of the
triplet \emph{vev}. Therefore we ignore them. 

We refer to the original Ref.~\cite{Csaki:2002qg} for the explicit
expression of the 21 EWPO in terms of the parameters $f$, $c$,
$c^{\prime}$, $x$ defined above (the list of the 21 EWPO can be found
in Sec.~\ref{sec:stat_exp}). We adapt their notation as follows:
\begin{equation}
	\Delta = \frac{v^2}{f^2}, \qquad \Delta^{\prime} =
        \frac{x^2}{16} \frac{v^2}{f^2}. 
\end{equation}

A derivation of the oblique parameters in the $L^2H$ model can be
found in Ref.~\cite{Kilian:2003xt}.

\subsubsection*{Littlest Higgs with \emph{T}-parity}

Due to the introduction of \emph{T}-parity, no \emph{T}-odd state can
contribute as external state at tree-level: therefore no contributions
to electroweak observables arise at tree-level from \emph{T}-odd
states. The only new particle which is \emph{T}-even is the
\emph{T}-even top partner $T_+$, but it can contribute at tree level
only to observables involving the SM top quark, such as its couplings
to $W$ and $Z$ bosons: since these couplings have not been measured
experimentally yet, no constraints arise at tree-level also from the
\emph{T}-even top partner. We will then consider only the one-loop
contributions to EWPO coming from the new \emph{T}-even/odd states,
using the results of
Ref.~\cite{Hubisz:2005tx,Berger:2012ec,Asano:2006nr}.  

At one loop, oblique corrections to the electroweak gauge boson
propagators induced by diagrams involving the top and its $T_+$
partner are given by 
\begin{eqnarray}
	S_{T_+} &=& \frac{s_{\be}^2}{2 \pi} \left[ \left( \frac{1}{3}
            - c_{\be}^2 \right) \log x_t + c_{\be}^2
          \frac{(1+x_t)^2}{(1-x_t)^2} + \frac{2 c_{\be}^2 x_t^2
            (3-x_t) \log x_t}{(1-x_t)^3} - \frac{8 c_{\be}^2}{3}
        \right] \nonumber \\ 
	T_{T_+} &=& \frac{3}{16 \pi} \frac{s_{\be}^2}{s_w^2 \, c_w^2}
        \frac{m_t^2}{m_Z^2} \left[ \frac{s_{\be}^2}{x_t} -1 -
          c_{\be}^2 - \frac{2 c_{\be}^2}{1-x_t} \log x_t
        \right] \label{TTeven} \\ 
	U_{T_+} &=& - \frac{s_{\be}^2}{2 \pi} \left[ s_{\be}^2 \log
          x_t + c_{\be}^2 \frac{(1+x_t)^2}{(1-x_t)^2} + \frac{2
            c_{\be}^2 x_t^2 (3- x_t) \log x_t}{(1-x_t)^3} - \frac{8
            c_{\be}^2}{3} \right] \nonumber  \qquad .
\end{eqnarray}
Here, $s_{\be}=\sin \be$ is the mixing angle in the right-handed top
sector, $x_t = m_t^2/m_{T+}^2$, and $s_w$ is the sine of the Weinberg
angle. 

The \emph{T}-odd top partner $T_{-}$ does not contribute to the
\emph{S,T,U} parameters since it is an $SU(2)_L$ singlet which does
not mix with the SM top. Moreover, the corrections from $T_-$ loops
are very small, and we do not include them in our fit. But the other
\emph{T}-odd heavy fermions coming from the interactions in
Eq. ($\ref{Todd}$) give a contribution to the \emph{T} parameter at
one-loop: under the assumption of flavor-independent $k$, the
contribution of each \emph{T}-odd fermion partner is given up to
$\mathcal{O} \left( v^2/f^2 \right)$ by 
\begin{equation}
	T_{\text{T-odd}} = - \frac{k^2}{192 \pi^2 \alpha_{w}} \frac{v^2}{f^2}.
\end{equation}
As explained in detail in the previous section, there is an upper
bound on the value of $k$ coming from four-fermion interactions
involving SM fields, Eq.~($\ref{kbound}$): the maximum contribution to
the \emph{T} parameter consistent with this bound becomes 
\begin{equation}
	|T_{\text{T-odd}}| \lesssim 0.05
\end{equation}
for each \emph{T}-odd fermion partner. The large number of
\emph{T}-odd partners (twelve) could thus have a sizable effect on
the EWPO: however a smaller value of $k$ could reduce this
contribution, and so we have not included it in our fit. 

Regarding the contribution from the gauge sector, the authors of
Ref.~\cite{Asano:2006nr} calculated that the total log-divergent
contribution due to the custodial $SU(2)$-violating tree-level mass
splitting of the \emph{T}-odd heavy $W_H^3$ and $W_H^{\pm}$ gauge
bosons completely vanishes, leaving only negligible finite terms of
order $v/f$ which we do not include in our analysis. 

Another important correction to both the \emph{S} and \emph{T}
parameters follows from the modified couplings of the Higgs boson to
the SM gauge bosons. In the SM, due to its renormalizability, the
one-loop contribution of the Higgs boson to the vector boson self
energies exactly cancels the logarithmic 
divergence arising from loops of would-be NGBs in the gauge-less
limit~\cite{Contino:2010rs}. As first noticed by the authors of
Ref.~\cite{Barbieri:2007bh}, the modified Higgs couplings to the SM
gauge bosons imply that the contribution of the Higgs to the
self-energy does not exactly cancel the infrared log-divergence
arising from the NGBs, leading to the following contributions to the
oblique parameters: 
\begin{eqnarray}
	S_h &=& - \frac{1}{6 \pi} \left( 1-y_W^2 \right) \log
        \frac{m_h}{\Lambda} \nonumber \\ 
	T_h &=& \frac{3}{8 \pi c_w^2} \left( 1-y_W^2 \right) \log
        \frac{m_h}{\Lambda}  \quad .
	\label{STh}
\end{eqnarray} 
$y_W$ parametrizes the shift of the coupling of one Higgs boson to the
SM gauge bosons in the usual notation, and $\Lambda = 4 \pi f$ is the
cut-off of the non-linear sigma model. In particular for the
\emph{LHT} model we obtain: 
\begin{eqnarray}
	S_h &=& - \frac{1}{18 \pi} \frac{v^2}{f^2} \log
        \frac{m_h}{\Lambda} \nonumber \\ 
	T_h &=& \frac{1}{8 \pi c_w^2} \frac{v^2}{f^2} \log
        \frac{m_h}{\Lambda}. 
\end{eqnarray} 

The \emph{LHT} model contains an additional \emph{T}-odd
$SU(2)_L$-triplet scalar field $\phi$: the effects on the \emph{S,T,U}
parameters are of order $\mathcal{O} ( v^4/m_{\phi}^4 )$ and therefore
negligible for $m_{\phi}$ in its natural range, around 1 TeV. We will
thus not include these effects in our fit. 

Finally, other possible contributions arise from new operators which
parametrize the effects of the UV physics on weak-scale
observables: 
\begin{eqnarray}
	S_{UV} &=& c_s \frac{4 m_W^2}{\pi g^2 f^2} \nonumber \\ 
	T_{UV} &=& - c_t \frac{m_W^2}{2 \pi e^2 g^2 f^2} \quad ,
\end{eqnarray}
where $c_s$ and $c_t$ are again coefficients of order one whose exact
values depend on the details of the UV physics, and which for
simplicity we assume to be equal to one as
in~\cite{Berger:2012ec}. All these different contributions to the
oblique parameters are then summed up. 

The only important non-oblique correction to the neutral-current
interactions which could affect the EWPO is the one-loop $T_+$
contribution to the $Z b_L \bar{b}_L$ vertex: to leading order in the
limit $m_{T+} \gg m_t \gg m_W$ it is given by 
\begin{equation}
	\delta \tilde{g}_L^{b \bar{b}} = \frac{g}{c_w} \frac{\alpha_{w}}{8 \pi s_w^2} \frac{m_t^4}{m_W^2 \, m_{T+}^2} R^2 \log \frac{m_{T+}^2}{m_t^2}.
\end{equation}
where we have used the notation of Appendix~\ref{app:ewpo} where more
details on the calculation of the EWPO can be found. 

With the explicit expressions of the oblique parameters \emph{S,T,U}
and of the neutral-current coefficient $\delta \tilde{g}_L^{b
  \bar{b}}$, we can finally obtain the explicit expressions of the 21
EWPO using the general results of Ref. \cite{Burgess:1993vc}
summarized in Appendix~\ref{app:ewpo}. The list of these variables and
their experimental values can be found in Sec.~\ref{sec:stat_exp}.

\subsubsection*{Simplest Little Higgs}

The dominant tree-level contributions to the oblique parameters in the
\emph{SLH} model come from the presence of a $Z^{\prime}$ boson with
$Z-Z^{\prime}$ mixing: their explicit expression can be found in
Ref.~\cite{Schmaltz:2004de,Marandella:2005wd} as 
\begin{equation}
	S_{Z'} = \frac{8 s_w^2}{\alpha_{w}}\frac{m_W^2}{g^2 f^2},
        \qquad T_{Z'} = \frac{1}{\alpha_{w}} \cdot \frac{v^2}{8 f^2}
        (1-t_w^2)^2. 
\end{equation}

The oblique parameters receive contributions also from the
modification in the Higgs couplings to the electroweak gauge bosons
w.r.t.~their SM values, cf.~Eq.~($\ref{STh}$): 
\begin{eqnarray}
	S_h &=& - \frac{1}{18 \pi} \frac{\tb^4-\tb^2+1}{\tb^2}
        \frac{v^2}{f^2} \log \frac{m_h}{\Lambda} \nonumber \\ 
	T_h &=& \frac{1}{8 \pi c_w^2} \frac{\tb^4-\tb^2+1}{\tb^2}
        \frac{v^2}{f^2} \log \frac{m_h}{\Lambda} \qquad .
\end{eqnarray}
Again, $\Lambda = 4 \pi f$ is the cut-off of the non-linear sigma
model. 

The corrections in the flavor sector can be read off from the
fermion-gauge interaction Lagrangians in \cite{Han:2005ru}: following
their assumptions, we will ignore right-handed mixing and choose the
Yukawa parameters in order to suppress the heavy-light mixing effects
in the first and second generations of quarks and in the
\emph{b}-quark sector\footnote{Using the notation of
  Ref.~\cite{Han:2005ru}, under these assumptions, we obtain the
  following relations: $\Delta u_i \simeq \Delta u_3 = V_{33}^{u *}
  \delta_{\nu} \tb^2 \frac{1-R^2}{R^2+\tb^2} \simeq \delta_{\nu} \tb^2
  \frac{1-R^2}{R^2+\tb^2}$, $\Delta Dd_j \simeq \Delta Dd \simeq
  \delta_{\nu}$ and $\Delta Sd_j \simeq \Delta Ss \simeq
  \delta_{\nu}$.}. Using the notation of Appendix~\ref{app:ewpo} and
defining the quantities
\begin{equation}
	\delta_{\nu} = -\frac{v}{\sqrt{2} f \tb}, \qquad \delta_Z = -
        \frac{(1-t_w^2) \sqrt{3-t_w^2}}{8 c_w} \frac{v^2}{f^2} \quad ,
\end{equation}
which parametrize the rotation to the mass eigenstates in the fermion-
and in the neutral gauge boson sectors, respectively, the corrections
to the charged-current couplings up to $\mathcal{O} \left( v^2/f^2
\right)$ are given in Table~$\ref{table:charged}$, while the
corrections to the neutral-current couplings (with $u \equiv u,c$ and
$d \equiv d,s$) are given in Table~$\ref{table:neutral}$. 
\begin{table}[!ht]
\centering
\begin{tabular}{|c||c|c|} \cline{2-3}
\multicolumn{1}{c|}{} & $\delta \tilde{h}_L$ & $\delta \tilde{h}_R$ \\ \hline
$\nu l$ & $-\delta_{\nu}^2/2$ & 0 \\ \hline
$ud$ & $-\delta_{\nu}^2/2$ & 0 \\ \hline
$cs$ & $-\delta_{\nu}^2/2$ & 0 \\ \hline
$tb$ & $-\frac{1}{2} \delta_{\nu}^2 \tb^4 \frac{(1-R^2)^2}{\left( R^2 +\tb^2 \right)^2}$ & 0 \\ \hline
\end{tabular}
\caption{\label{table:charged}Corrections of the charged-current
  couplings in \emph{SLH}.} 
\end{table}%
\begin{table}[!ht]
\centering
\begin{tabular}{|c||c|c|} \cline{2-3}
\multicolumn{1}{c|}{} & $\delta \tilde{g}_L$ & $\delta \tilde{g}_R$ \\ \hline
$\nu \nu$ & $-\delta_{\nu}^2/2 + (1/2-s_w^2) \delta_Z/\sqrt{3-4 s_w^2}$ & 0 \\ \hline
$ll$ & $(1/2-s_w^2) \delta_Z/\sqrt{3-4 s_w^2}$ & $s_w^2 \delta_Z/\sqrt{3-4 s_w^2}$ \\ \hline
$uu$ & $(-1/2+2/3 s_w^2) \delta_Z/\sqrt{3-4 s_w^2}$ & $-2/3 s_w^2 \delta_Z/\sqrt{3-4 s_w^2}$ \\ \hline
$tt$ & $-\frac{1}{2} \delta_{\nu}^2 \tb^4 \frac{(1-R^2)^2}{\left( R^2 +\tb^2 \right)^2} + (1/2-1/3 s_w^2) \delta_Z/\sqrt{3-4 s_w^2}$ & $-2/3 s_w^2 \delta_Z/\sqrt{3-4 s_w^2}$ \\ \hline
$dd$ & $\delta_{\nu}^2/2 + (-1/2+2/3 s_w^2) \delta_Z/\sqrt{3-4 s_w^2}$ & $1/3 s_w^2 \delta_Z/\sqrt{3-4 s_w^2}$ \\ \hline
$bb$ & $(1/2-1/3 s_w^2) \delta_Z/\sqrt{3-4 s_w^2}$ & $1/3 s_w^2 \delta_Z/\sqrt{3-4 s_w^2}$ \\ \hline
\end{tabular}
\caption{\label{table:neutral}Corrections of the neutral-current
  couplings in \emph{SLH}.} 
\end{table}%

With the explicit expressions of the oblique parameters \emph{S,T,U}
and of the charged- and neutral-current coefficients, we can finally
obtain the explicit expressions of the 21 EWPO using the
general results of Ref. \cite{Burgess:1993vc} cited in
Appendix~\ref{app:ewpo}. 


\section{Statistical Method and Experimental Data}
\label{sec:stat_exp}

It is customary for the experimental collaborations to express the
results of the SM-like Higgs searches in terms of a \emph{signal
  strength modifier} $\mu$, defined as the factor by which the SM
Higgs signal is modified for a given value of $m_{h}$: 
\begin{equation}
	\mu^{i} = \frac{ n_{S}^{i} }{ n^{SM, \, i}_{S} } =
        \frac{\sum_{p} \sigma_{p} \cdot \epsilon_{i}^{p}}{\sum_{p}
          \sigma_{p}^{SM} \cdot \epsilon_{i}^{p}} \cdot
        \frac{BR_{i}}{BR_{i}^{SM}} 
	\label{mu}
\end{equation}
where $i$, $p$ are indices for a specific decay channel and production
mode, respectively. $\epsilon_{i}^{p}$ is the efficiency of the
kinematic cuts for a given production mode $p$ and decay channel $i$,
and $n_{S}^i$ is the number of expected Higgs signal events evaluated
in a chosen model (e.g.~$n^{SM, \, i}_{S}$ is evaluated in the SM). 

For each Higgs decay channel considered, the ATLAS and CMS
collaborations usually report the 95$\%$ \emph{CL} limit on $\mu$
($\mu_{95\%}$) and the best-fit value $\hat{\mu}$ for a given
hypothesis on $m_h$. In particular, values $\mu_{95\%} < 1$ exclude at
95$\%$ \emph{CL} the SM Higgs for that particular value of the Higgs
mass. The efficiencies of the kinematic cuts are instead not reported
by the collaborations (the only exceptions are all the $\gamma \gamma$
and CMS $\tau \tau$ 8 TeV channels), making it thus very hard (if not
impossible) to correctly compare a theory prediction with the observed
data.  

To implement a $\chi^2$ analysis we follow the procedure described
in~\cite{Espinosa:2012ir}. One defines the covariance matrix
$\mathcal{C}$ of the observables, and $\Delta \theta_i$ as the vector
of the difference in the observed and predicted value of the
observables, which is a function of the free parameters of the
model. The $\chi^2$ measure is then given by 
\begin{equation}
	\chi^2 = \left( \Delta \theta_i \right)^T \left( \mathcal{C}^{-1} \right)_{ij} \left( \Delta \theta_j \right).
	\label{chi2}
\end{equation}
The 95$\%$ and 99$\%$ best-fit \emph{CL} regions are then defined
by the cumulative distribution function for an appropriate number of
degrees of freedom (\emph{d.o.f.}). 

\subsection{Higgs searches}

First, we consider as observables for the $\chi^2$ measure
($\ref{chi2}$) the different best-fit values of the signal strength
modifiers of all available public data reported by the ATLAS and CMS
collaborations for the 7 and 8 TeV Higgs searches. In particular, we
include in our analysis the 7 TeV $\sim 5$ fb$^{-1}$ and 8 TeV
$\sim 6$ fb$^{-1}$ data from the July 2012 publications of both
collaborations, and also the latest December 2012 update of up to
$\sim 13$ fb$^{-1}$ of many of the 8 TeV samples. 

For our analysis, we have taken the matrix $\mathcal{C}$ to be diagonal
with the sum of the square of the 1$\sigma$ theory and experimental
errors as diagonal entries: off-diagonal correlation coefficients
are indeed neglected, as correlation coefficients are currently not
supplied by the experimental collaborations. As already discussed
in~\cite{Azatov:2012qz}, the absence of information regarding 
correlations in fits of Higgs couplings is not a significant
limitation, given the current level of statistical uncertainty. The
authors in~\cite{Azatov:2012qz} claim indeed that the error on the
best fit point assuming zero correlation is less than 1$\%$, at least
in the CMS $\gamma \gamma$ final state. 

For the experimental errors we use the quoted 1$\sigma$ errors on the
reported signal strength ($\delta \mu_{i,\text{exp}}$), while for
theoretical uncertainties we propagate the cross section error $\delta
\sigma_i$ as an uncertainty on the signal strength modifier: 
\begin{equation}
	\delta \mu_{i,\text{th}} = \mu_i \left( \frac{\sqrt{\sum_j
              r_j^2 \cdot \delta \sigma_j^2}}{\sum r_j \cdot \sigma_j}
          - \frac{\sqrt{\sum_j \delta \sigma_j^2}}{\sum_j \sigma_j}
        \right)  \qquad ,
\end{equation}
where $r_j$ is the appropriate rescaling factor for each cross section
$j$. In this way, the $\chi^2$ measure reduces to the usual form 
\begin{equation}
	\chi^2 = \sum_i \frac{(\mu_i - \hat{\mu}_i)^2}{\sigma_i^2}
\end{equation}
where $\mu_i$ is the $i$-th signal strength predicted by the model as
a function of the free parameters, $\hat{\mu}_i$ is the respective
best-fit value, and $\sigma_i = \sqrt{\delta
  \mu_{i,\text{exp}}^2+\delta \mu_{i,\text{th}}^2}$ the total
uncertainty. 

We summarize in Table~$\ref{table:bestfit}$ the available data,
reporting in particular the different best-fit values $\hat{\mu}_i$
and the reference masses at which the single $\hat{\mu}_i$ are
evaluated. Notice that these reference masses are not necessarily 
the masses at which the highest local significance has been 
obtained for each channel. We have chosen the best-fit values of the 
signal strengths and the corresponding masses in order to be able to 
reconstruct separately (for which only 7 TeV and combined results are
given by the experiments) the 7 and 8 TeV signal strengths. For
example, even if the highest significance in the 7+8 TeV combined
analysis of the ATLAS $ZZ$ channel~\cite{ATLASDecZZ} has been found
for a Higgs mass of 123.5 GeV ($\hat{\mu} = 1.3 \pm 0.5$), we used the
7+8 TeV signal strength for a Higgs mass of 126 GeV ($\hat{\mu} = 0.8
\pm 0.4$) for reconstructing the 8 TeV signal strength reported
in Table~$\ref{table:bestfit}$, in order to comply with the 7 TeV
signal strength which has been given in Ref.~\cite{:2012gk} for
$m_{h}=126$ GeV.  

\begin{table}[!ht]
\begin{center}
\begin{tabular}[c]{|c||c|c|c|c|}
\hline
\textbf{ATLAS 7 TeV} &  $m_h$ [GeV] & $\hat{\mu}$ \\ \hline \hline
$\gamma \gamma_{\text{\tiny{UClPT}}}$ \cite{ATLASgamma} & 126.5 & 0.5 $\pm$ 1.5 \\ \hline
$\gamma \gamma_{\text{\tiny{UChPT}}}$ \cite{ATLASgamma} & 126.5 & 0.2 $\pm$ 2.0 \\ \hline
$\gamma \gamma_{\text{\tiny{URlPT}}}$ \cite{ATLASgamma} & 126.5 & 2.5 $\pm$ 1.7 \\ \hline
$\gamma \gamma_{\text{\tiny{URhPT}}}$ \cite{ATLASgamma} & 126.5 & 10.4 $\pm$ 3.7 \\ \hline
$\gamma \gamma_{\text{\tiny{CClPT}}}$ \cite{ATLASgamma} & 126.5 & 6.1 $\pm$ 2.6 \\ \hline
$\gamma \gamma_{\text{\tiny{CChPT}}}$ \cite{ATLASgamma} & 126.5 & -4.4 $\pm$ 1.8 \\ \hline
$\gamma \gamma_{\text{\tiny{CRlPT}}}$ \cite{ATLASgamma} & 126.5 & 2.7 $\pm$ 2.0 \\ \hline
$\gamma \gamma_{\text{\tiny{CRhPT}}}$ \cite{ATLASgamma} & 126.5 & -1.6 $\pm$ 2.9 \\ \hline
$\gamma \gamma_{\text{\tiny{CT}}}$ \cite{ATLASgamma} & 126.5 & 0.3 $\pm$ 3.6 \\ \hline
$\gamma \gamma_{\text{\tiny{jj}}}$ \cite{ATLASgamma} & 126.5 & 2.7 $\pm$ 1.9 \\ \hline \hline
$ZZ$ \cite{:2012gk} & 126.0 & 1.4 $\pm$ 1.1 \\ \hline
$WW$ \cite{:2012gk} & 126.0 & 0.5 $\pm$ 0.6 \\ \hline
$bb$ \cite{ATLASNovbb} & 125.0 & -2.7 $\pm$ 1.6 \\ \hline
$\tau \tau$ \cite{:2012gk} & 126.0 & 0.4 $\pm$ 1.8 \\ \hline
\end{tabular}
\hspace{2pt}
\begin{tabular}[c]{|c||c|c|c|c|}
\hline
\textbf{ATLAS 8 TeV} & $m_h$ [GeV] & $\hat{\mu}$ \\ \hline \hline
$\gamma \gamma_{\text{\tiny{UClPT}}}$ \cite{ATLASDecgamma} & 126.5 & 1.0 $\pm$ 0.9 \\ \hline
$\gamma \gamma_{\text{\tiny{UChPT}}}$ \cite{ATLASDecgamma} & 126.5 & 0.3 $\pm$ 1.2 \\ \hline
$\gamma \gamma_{\text{\tiny{URlPT}}}$ \cite{ATLASDecgamma} & 126.5 & 2.9 $\pm$ 1.2 \\ \hline
$\gamma \gamma_{\text{\tiny{URhPT}}}$ \cite{ATLASDecgamma} & 126.5 & 1.8 $\pm$ 1.4 \\ \hline
$\gamma \gamma_{\text{\tiny{CClPT}}}$ \cite{ATLASDecgamma} & 126.5 & 1.5 $\pm$ 1.2 \\ \hline
$\gamma \gamma_{\text{\tiny{CChPT}}}$ \cite{ATLASDecgamma} & 126.5 & 1.0 $\pm$ 1.6 \\ \hline
$\gamma \gamma_{\text{\tiny{CRlPT}}}$ \cite{ATLASDecgamma} & 126.5 & 2.3 $\pm$ 1.2 \\ \hline
$\gamma \gamma_{\text{\tiny{CRhPT}}}$ \cite{ATLASDecgamma} & 126.5 & 0.5 $\pm$ 1.6 \\ \hline
$\gamma \gamma_{\text{\tiny{CT}}}$ \cite{ATLASDecgamma} & 126.5 & 2.0 $\pm$ 2.0 \\ \hline
$\gamma \gamma_{\text{\tiny{2jhm}}}$ \cite{ATLASDecgamma} & 126.5 & 2.0 $\pm$ 1.1 \\ \hline
$\gamma \gamma_{\text{\tiny{2jlm}}}$ \cite{ATLASDecgamma} & 126.5 & 3.6 $\pm$ 2.1 \\ \hline
$\gamma \gamma_{\text{\tiny{LT}}}$ \cite{ATLASDecgamma} & 126.5 & 1.2 $\pm$ 2.2 \\ \hline \hline
$ZZ$ \cite{ATLASDecZZ}\textcolor{red}{*} & 126.0 & 0.7 $\pm$ 0.4 \\ \hline
$WW$ \cite{ATLASNov} & 126.0 & 1.4 $\pm$ 0.6 \\ \hline
$bb$ \cite{ATLASNovbb} & 125.0 & 1.0 $\pm$ 1.4 \\ \hline
$\tau \tau$ \cite{ATLASNov}\textcolor{red}{*} & 126.0 & 0.7 $\pm$ 0.8 \\ \hline
\end{tabular}
\end{center}

\begin{center}
\begin{tabular}[c]{|c||c|c|c|c|}
\hline
\textbf{CMS 7 TeV} & $m_h$ [GeV] & $\hat{\mu}$ \\ \hline \hline
$\gamma \gamma_{\text{\tiny{cat0}}}$ \cite{CMSgamma} & 125.0 & 3.2 $\pm$ 1.8 \\ \hline
$\gamma \gamma_{\text{\tiny{cat1}}}$ \cite{CMSgamma} & 125.0 & 0.7 $\pm$ 0.9 \\ \hline
$\gamma \gamma_{\text{\tiny{cat2}}}$ \cite{CMSgamma} & 125.0 & 0.7 $\pm$ 1.2 \\ \hline
$\gamma \gamma_{\text{\tiny{cat3}}}$ \cite{CMSgamma} & 125.0 & 1.5 $\pm$ 1.6 \\ \hline
$\gamma \gamma_{\text{\tiny{jj}}}$ \cite{CMSgamma} & 125.0 & 4.2 $\pm$ 2.1 \\ \hline \hline
$ZZ$ \cite{CMSJuly} & 125.0 & 0.6 $\pm$ 0.6 \\ \hline
$WW$ \cite{CMSJuly} & 125.0 & 0.4 $\pm$ 0.6 \\ \hline
$bb$ \cite{CMSJuly} & 125.0 & 0.6 $\pm$ 1.2 \\ \hline
$\tau \tau$ \cite{CMSNovtau} & 125.0 & 1.0 $\pm$ 0.9 \\ \hline
\end{tabular}
\hspace{2pt}
\begin{tabular}[c]{|c||c|c|c|c|c|}
\hline
\textbf{CMS 8 TeV} & $m_h$ [GeV] & $\hat{\mu}$ \\ \hline \hline
$\gamma \gamma_{\text{\tiny{cat0}}}$ \cite{CMSgamma} & 125.0 & 1.4 $\pm$ 1.2 \\ \hline
$\gamma \gamma_{\text{\tiny{cat1}}}$ \cite{CMSgamma} & 125.0 & 1.5 $\pm$ 1.0 \\ \hline
$\gamma \gamma_{\text{\tiny{cat2}}}$ \cite{CMSgamma} & 125.0 & 0.9 $\pm$ 1.2 \\ \hline
$\gamma \gamma_{\text{\tiny{cat3}}}$ \cite{CMSgamma} & 125.0 & 3.8 $\pm$ 1.8 \\ \hline
$\gamma \gamma_{\text{\tiny{jj loose}}}$ \cite{CMSgamma} & 125.0 & -0.6 $\pm$ 2.0 \\ \hline
$\gamma \gamma_{\text{\tiny{jj tight}}}$ \cite{CMSgamma} & 125.0 & 1.3 $\pm$ 1.6 \\ \hline \hline
$ZZ$ \cite{CMSNovZZ}\textcolor{red}{*} & 125.0 & 0.9 $\pm$ 0.4 \\ \hline
$WW$ \cite{CMSNovWW}\textcolor{red}{*} & 125.0 & 0.8 $\pm$ 0.3 \\ \hline
$bb$ \cite{CMSNovbb}\textcolor{red}{*} & 125.0 & 1.5 $\pm$ 0.7 \\ \hline
$\tau \tau$ \cite{CMSNovtau} & 125.0 & 0.6 $\pm$ 0.6 \\ \hline
\end{tabular}
\end{center}
\caption{Signal-strength best-fit values of the $7$ and $8$ TeV
  samples collected by ATLAS and CMS. The red asterisks mark the 8 TeV
  channels for which the best-fit signal-strengths have been
  reconstructed from the 7 and 7+8 TeV values.}  
\label{table:bestfit}
\end{table}%

For some channels, indeed only the combined 7+8 TeV signal strengths
are available in addition to the 7 TeV results. As suggested
in~\cite{Espinosa:2012ir}, one can assume a Gaussian approximation for 
the probability density functions (\emph{pdf}) describing the
different signal strengths, i.e. 
\begin{equation}
	p[\mu_i | \hat{\mu}_i, \sigma_i] \simeq e^{(\mu_i - \hat{\mu}_i)^2/(2 \sigma_i^2)}
\end{equation}
and obtain the combined \emph{pdf} by multiplying the individual
channel \emph{pdf}s. The combined \emph{pdf} is therefore also
Gaussian, with central value $\hat{\mu}_c$ and width $\sigma_c$
approximately given by 
\begin{equation}
	\frac{1}{\sigma_c^2} = \sum_i \frac{1}{\sigma_i^2}, \qquad
        \frac{\hat{\mu}_c}{\sigma_c^2} = \sum_i
        \frac{\hat{\mu}_i}{\sigma_i^2}. 
\end{equation} 
By solving the previous equations, one can then reconstruct the
unknown 8 TeV data from the reported 7 and 7+8 TeV data, as has
been done for the channels marked with a red asterisk in
Table~$\ref{table:bestfit}$. 

It is to be noted that the collaborations have reported the best-fit
values of the 7 and 8 TeV diphoton channels exclusively with respect
to the different selection cut categories, and reported also the
different cut efficiencies for the single subchannels,
cf. Ref.~\cite{ATLASgamma,CMSgamma}. Therefore we were able to add all
single diphoton subchannel contributions to the $\chi^2$ measure, 
exclusively with respect to the production modes.  

For the $h \rightarrow bb$ channels we assume the results to be fully
dominated by the Higgs-Strahlung production mode, neglecting the
contributions from the other production modes. For all other channels
we considered the signal as inclusive with respect to the production
modes, neglecting thus the cut efficiencies, since they have not been
reported. 

The SM parameters have been obtained from the updated values of the
Particle Data Group Collaboration~\cite{Beringer:1900zz}, while for
the SM Higgs production cross sections with respective uncertainties
and branching ratios we have used the recommended values by the
\emph{LHC Higgs Cross Section Working Group}~\cite{Dittmaier:2011ti}.

\subsection{Electroweak Precision Data}

We incorporate EWPO by directly adding their contribution to the
$\chi^2$ measure. In particular, we include the contribution from the
following 21 different low-energy and $Z$-pole precision observables
for $m_h=124.5$ GeV~\cite{Beringer:1900zz}, as summarized in Table
$\ref{table:EWPO}$. 
\begin{table}[!ht]
\begin{center}
\begin{tabular}{|c||c|c|} \cline{2-3}
\multicolumn{1}{c|}{} & Value & SM prediction \\ \hline \hline
$\Gamma_Z$ [GeV] & 2.4952 $\pm$ 0.0023 & 2.4961 $\pm$ 0.0010 \\ \hline
$R_{e}$ & 20.804 $\pm$ 0.050 & 20.744 $\pm$ 0.011 \\ \hline
$R_{\mu}$ & 20.785 $\pm$ 0.033 & 20.744 $\pm$ 0.011 \\ \hline
$R_{\tau}$ & 20.764 $\pm$ 0.045 & 20.789 $\pm$ 0.011 \\ \hline
$\sigma_{\text{had}}$ [nb] & 41.541 $\pm$ 0.037 & 41.477 $\pm$ 0.009 \\ \hline
$R_{b}$ & 0.21629 $\pm$ 0.00066 & 0.21576 $\pm$ 0.00004 \\ \hline
$R_{c}$ & 0.1721 $\pm$ 0.0030 & 0.17227 $\pm$ 0.00004 \\ \hline
$A_{\text{FB}}^{e}$ & 0.0145 $\pm$ 0.0025 & 0.01633 $\pm$ 0.00021 \\ \hline
$A_{\text{FB}}^{\mu}$ & 0.0169 $\pm$ 0.0013 & 0.01633 $\pm$ 0.00021 \\ \hline
$A_{\text{FB}}^{\tau}$ & 0.0188 $\pm$ 0.0017 & 0.01633 $\pm$ 0.00021 \\ \hline
$A_{\tau}(P_{\tau})$ & 0.1439 $\pm$ 0.0043 & 0.1475 $\pm$ 0.0010 \\ \hline
$A_{e}(P_{\tau})$ & 0.1498 $\pm$ 0.0049 & 0.1475 $\pm$ 0.0010 \\ \hline
$A_{\text{FB}}^{b}$ & 0.0992 $\pm$ 0.0016 & 0.1034 $\pm$ 0.0007 \\ \hline
$A_{\text{FB}}^{c}$ & 0.0707 $\pm$ 0.0035 & 0.0739 $\pm$ 0.0005 \\ \hline
$A_{\text{LR}}$ & 0.15138 $\pm$ 0.00216 & 0.1475 $\pm$ 0.0010 \\ \hline
$m_W$ [GeV] & 80.420 $\pm$ 0.031 & 80.381 $\pm$ 0.014 \\ \hline
$g_L^2$ & 0.3009 $\pm$ 0.0028 & 0.3040 $\pm$ 0.0002 \\ \hline
$g_R^2$ & 0.0328 $\pm$ 0.0030 & 0.03001 $\pm$ 0.00002 \\ \hline
$g_V^{\nu e}$ & -0.040 $\pm$ 0.015 & -0.0398 $\pm$ 0.0003 \\ \hline
$g_A^{\nu e}$ & -0.507 $\pm$ 0.014 & -0.5064 $\pm$ 0.0001 \\ \hline
$Q_W(Cs)$ & -73.20 $\pm$ 0.35 & -73.23 $\pm$ 0.02 \\ \hline
\end{tabular}
\end{center}
\caption{\label{table:EWPO} Experimental values and SM predictions of
  the 21 different EWPO.} 
\end{table}

Since no correlation coefficients are supplied for these 21 observables, we will assume them as independent and add their contribution to the $\chi^2$ measure as
\begin{equation}
	\chi^2 = \sum_{i} \frac{( \mathcal{O}_i-\hat{\mathcal{O}}_i
          )^2}{\sigma_i^2} \qquad .
\end{equation}
Here, $\mathcal{O}_i$ is the $i$-th observable predicted by the model
as a function of the free parameters, $\hat{\mathcal{O}}_i$ is the
respective measured value, and $\sigma_i$ the experimental uncertainty.  


\section{Results}
\label{sec:results}

In order to calculate the updated exclusion contours for the different
models, we need to determine the explicit expression of the signal
strength modifier 
$\mu^{i}$ 
\begin{equation}
	\mu^{i}  = \frac{\sum_{p} \sigma_{p}^{\text{LH}} \cdot
          \epsilon_{i}^{p}}{\sum_{p} \sigma_{p}^{\text{SM}} \cdot
          \epsilon_{i}^{p} } \cdot \frac{BR_{i}^{\text{LH}}
        }{BR_{i}^{SM}} 
	\label{signalmod}
\end{equation}
for each decay channel $i$. $\mu^i$ depends on the different free
parameters of the model under which it is evaluated: in particular,
the three models we are considering share a free parameter, namely the
dimensionless ratio $v_{SM}/f$ with $v_{SM}= 246$ GeV, where $f$ is
the spontaneous symmetry breaking scale of the respective global
symmetries. The ratio $v_{SM}/f$ varies in the interval $[0,1]$: the
lower bound is the \emph{SM-} or \emph{decoupling-limit} where all the
modifications due to the Little Higgs structure are vanishing,
recovering the SM results, while the upper limit is set in order to
constrain the global symmetry breaking scale to be greater than the
EWSB scale. The other free parameters are model-dependent, and under
few assumptions we will consider only one extra free parameter for
each of the three models.  

In the \emph{L$^2$H} model, the other free parameters are the mixing
angles $c$, $c^{\prime}$ in the gauge sector ($\ref{cdef}$), the ratio
$R = \lambda_1 / \lambda_2$ of the couplings in the top sector
($\ref{RdefLH}$), and the parameter $x$ proportional to the triplet
\emph{vev} ($\ref{xdef}$). However, we will fix the value of $R$ to a
reference value of one, since our results will be with good
approximation independent on the particular value of $R$, as a
consequence of the collective symmetry breaking mechanism, as we will
show later. For the remaining free parameters, we will let only the
mixing angle $c$ vary between $[0.1,0.995]$, while presenting our
results for few different choices of $x$ and $c^{\prime}$. 

In the \emph{LHT} model, besides the scale $f$, the only other free
parameter for our study is again the ratio $R = \lambda_1 / \lambda_2$
of the couplings in the \emph{T}-even top sector ($\ref{topLHT}$). In
Ref. \cite{Belyaev:2006jh} the authors have performed a study to fix
the allowed range for $R$ in \emph{LHT}: they obtained $R \lesssim
3.3$ by calculating the $J=1$ partial-wave amplitudes in the coupled
system of ($t\bar{t}$, $T\bar{T}_{+}$, $b \bar{b}$, $WW$, $Z h$)
states to estimate the tree-level unitarity limit of the corresponding
scattering amplitudes. Therefore we will vary  $R$ between
$[0.1,3.3]$, where the lower limit is na\"{\i}vely chosen by
naturalness arguments.  

In the \emph{SLH} model the free parameters $R = \lambda_1^t /
\lambda_2^t$ and $\mu_{\phi}^2$ are fixed by the requirement of EWSB,
as described in Sec.~\ref{sec:frame} , leaving $f$ and $t_{\beta}$ as free
parameters of our study, where $t_{\beta}$ is the ratio of the
\emph{vev}s of the two scalar fields $\phi_{1,2}$
($\ref{scalarsSLH}$). We will let $t_{\be}$ vary between
$[1.0,15]$, where both limits are again na\"{\i}vely chosen by
naturalness. We will also require the perturbative constraint
($\ref{perturbative}$) to be satisfied: this will restrict the allowed
values of $t_{\be}$ for a given value of $f$. 

Following the procedure of Sec.~\ref{sec:stat_exp}, we determine the $\chi^2$
measure ($\ref{chi2}$) including the contributions from the
deviations among the predicted and reported best-fit values of the
different signal strength modifiers, and from the electroweak
observables. The $95\%$ and $99\%$ \emph{CL} allowed regions are then
defined by the cumulative distribution function for an appropriate
number of d.o.f.: having a total of 49 different best-fit channels, 
and 21 EWPO, the total number of d.o.f. is 70, 
since no free parameters have been fitted to the data.  

\begin{figure}[!ht]
\centering
\includegraphics[width=0.483\textwidth]{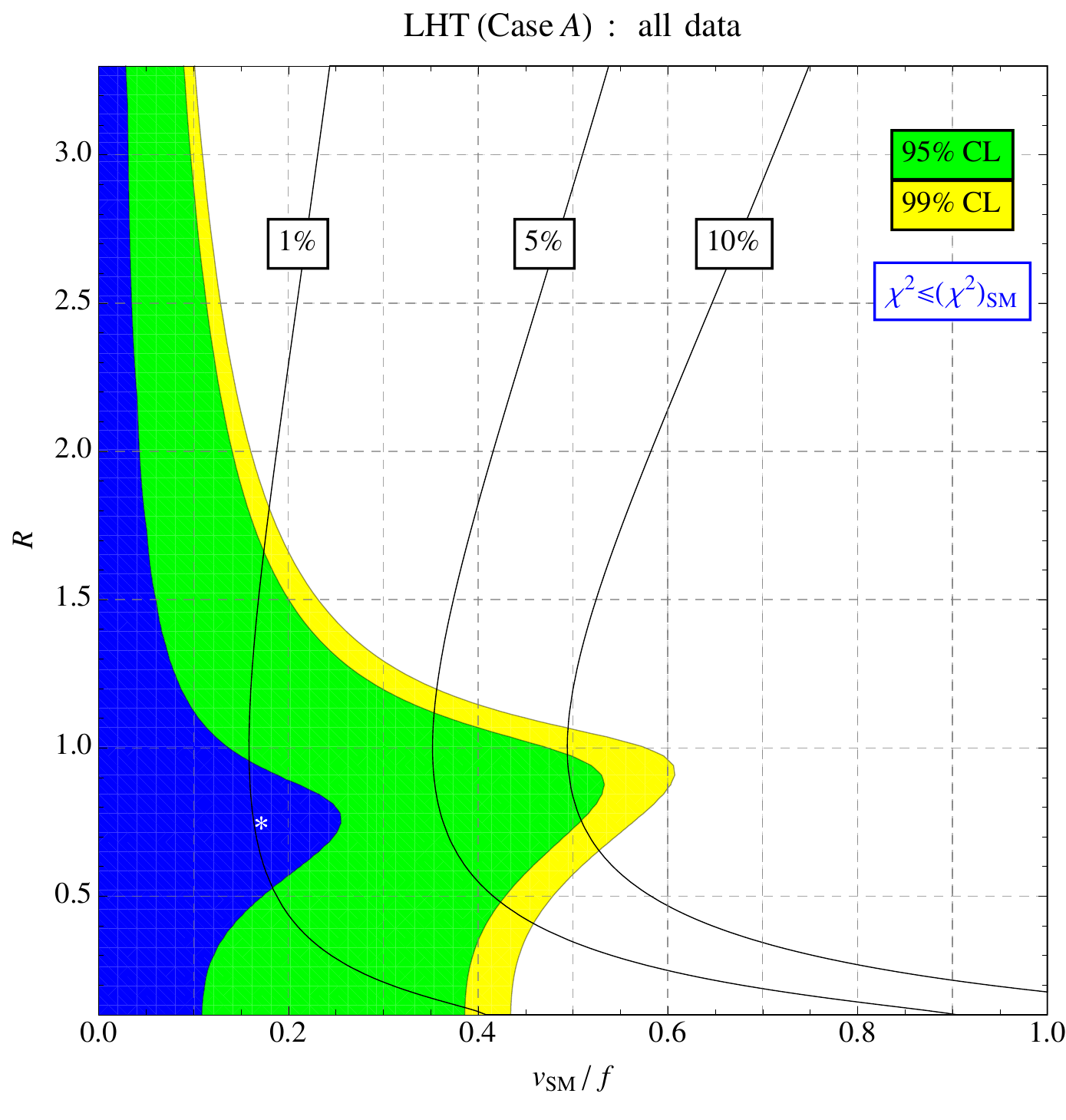} \quad
\includegraphics[width=0.483\textwidth]{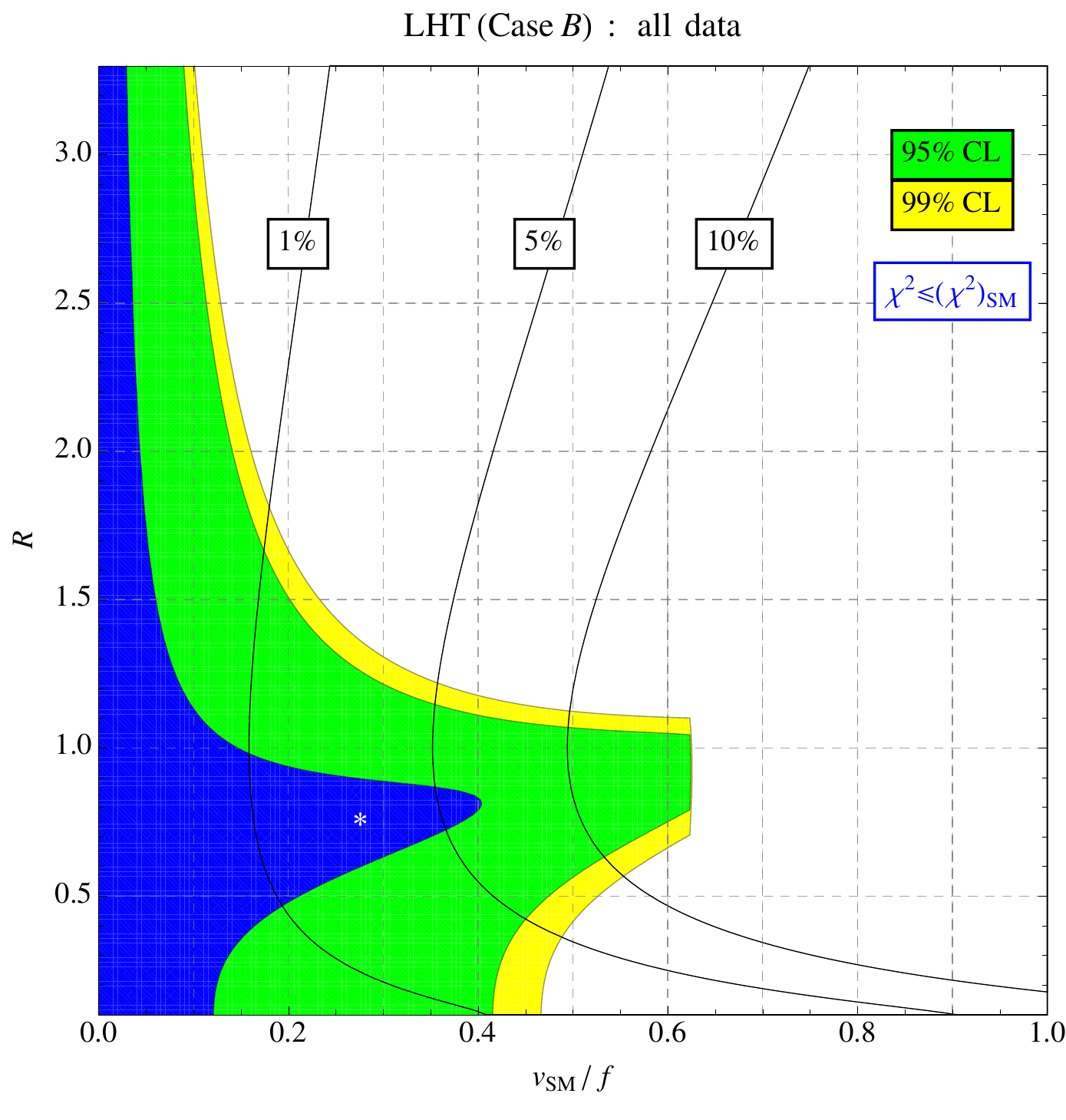}
\includegraphics[width=0.483\textwidth]{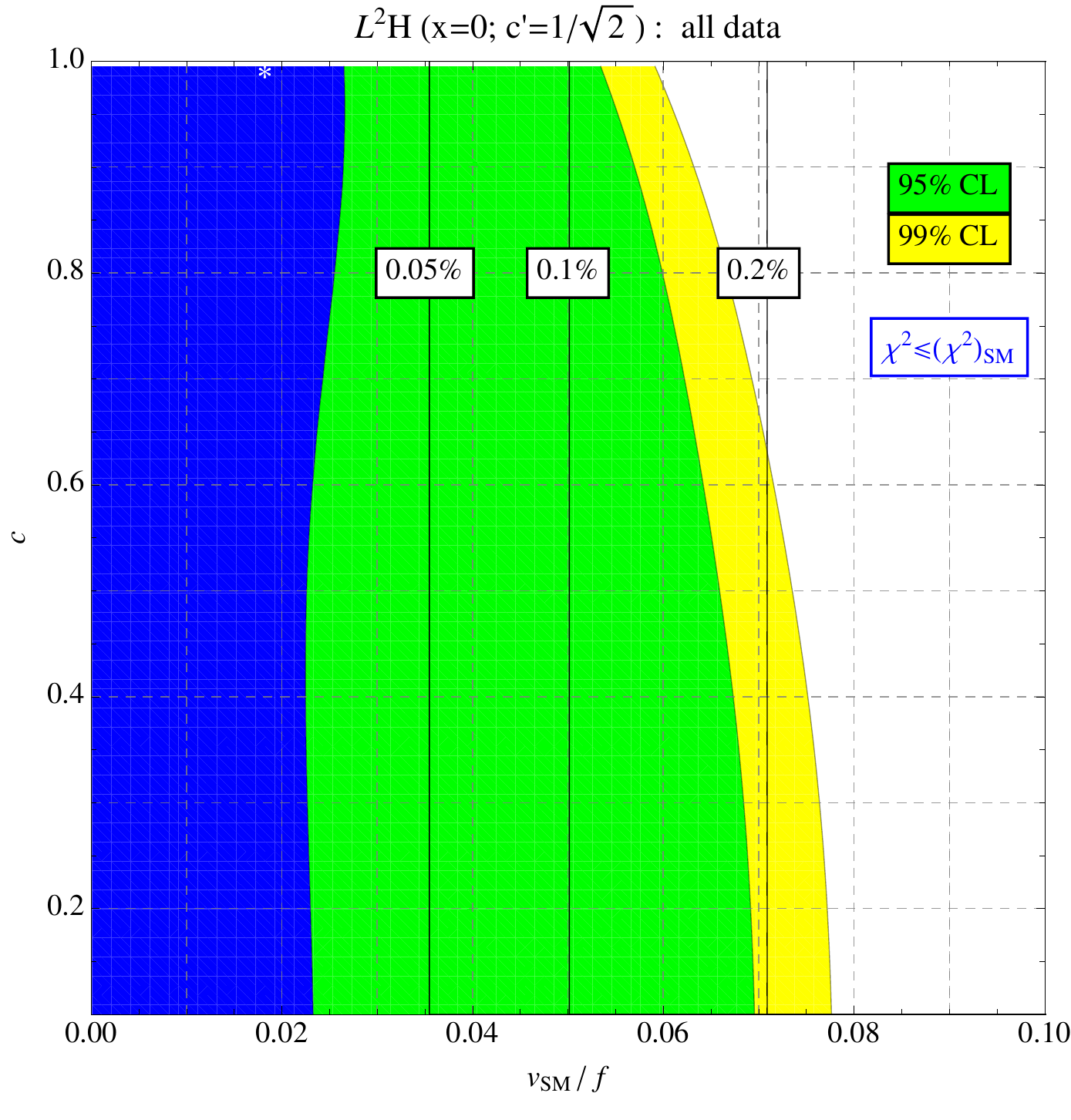} \quad
\includegraphics[width=0.483\textwidth]{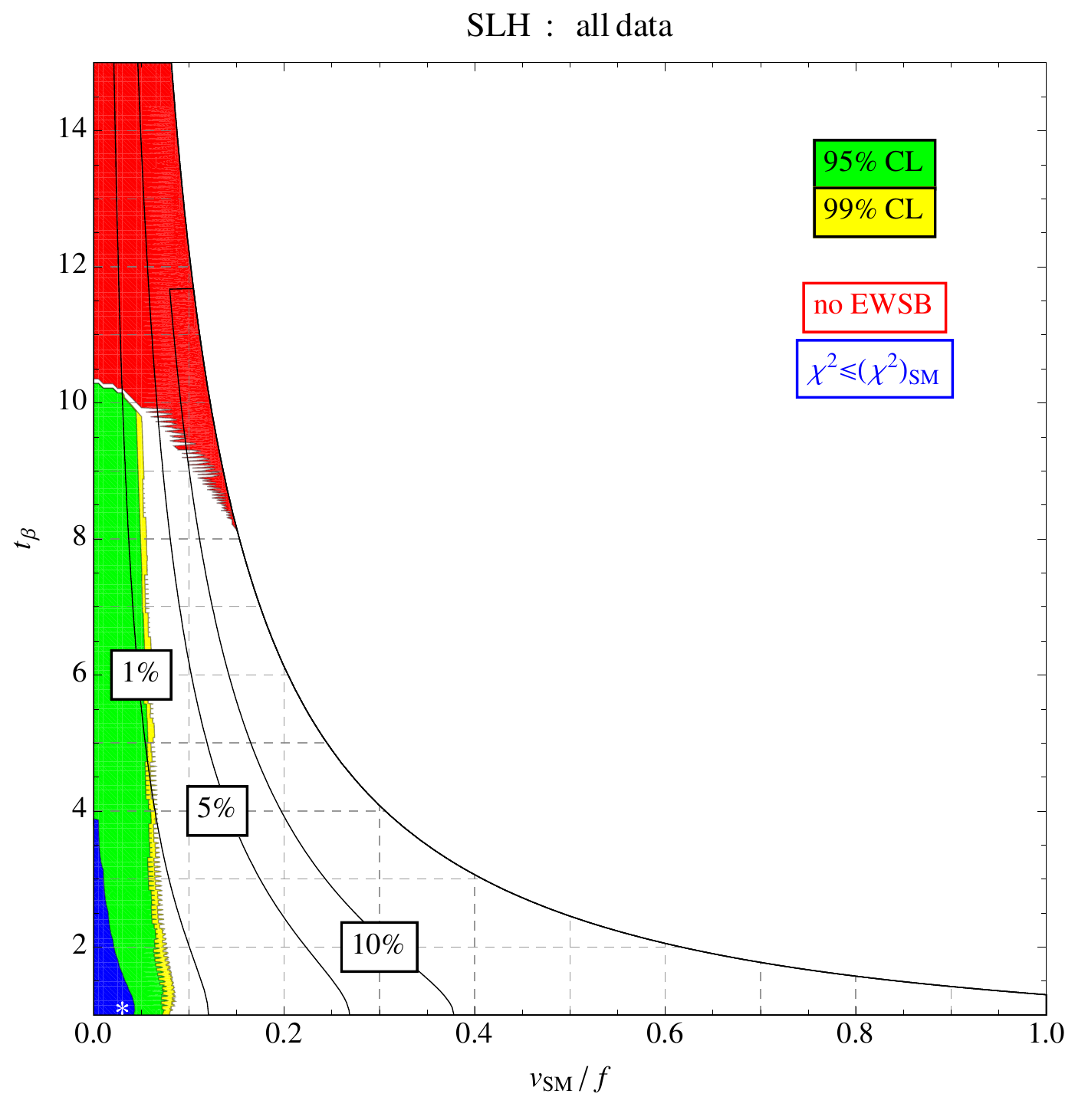}
\caption{Allowed contours at 95$\%$ and 99$\%$ \emph{CL} considering
  the whole available dataset for \emph{LHT} \emph{Case A} (up left),
  \emph{LHT} \emph{Case B} (up right), \emph{L$^2$H} (down left) and
  \emph{SLH} (down right). In the red region, no EWSB is
  possible. In the blue regions we found a lower $\chi^{2}$ than
  the SM $\chi^{2}$: the white points have the minimum $\chi^2$. The
  thick black lines represent contours of required fine-tuning.}  
\label{chi2fit}
\end{figure}

We present in Fig.~\ref{chi2fit} the updated exclusion contours for
the different models considered, distinguishing in particular Case
\emph{A} and Case 
\emph{B} of the \emph{LHT}. As the \emph{L$^2$H} model is concerned,
for the relative plot we have fixed $x=0$ and $c^{\prime}=1/\sqrt{2}$,
and restrict ourself to the region $v/f \in [0,0.1]$ where the EWPO
are satisfied and no tree-level decay of the Higgs involving new heavy
partners is kinematically allowed, as already mentioned in
Sec.~\ref{sec:frame}. In the red region of the \emph{SLH} plot the
EWSB cannot be realized, and this region is therefore excluded. In the
blue region of each plot we registered a total $\chi^{2}$ which is
lower than the $\chi^2$ of the SM, with all signal strength modifiers
set to 1 and all EWPO to their SM predictions. The minimum of the
$\chi^2$-value is denoted by the white points in the plots. The black
lines represent contours of required fine-tuning inside the model
setup, as we will explain later. 

The new lower bounds of the symmetry breaking scale $f$ at 99$\%$ and
95$\%$ \emph{CL} within each model, as well as the value of $f$ at
which the minimum $\chi^{2}$ has been determined, are summarized in
Table $\ref{table:lowerval}$. 
\begin{table}[!ht]
\begin{center}
\begin{tabular}{|c|c|c|c|c|} \cline{2-5}
\multicolumn{1}{c|}{} & \textbf{\emph{LHT} Case \emph{A}} &
\textbf{\emph{LHT} Case \emph{B}} & \textbf{\emph{L$^2$H}} &
\textbf{\emph{SLH}} \\ \hline 
$f^{99\%}_{\text{min}}$ [TeV] & 0.41 & 0.39 & 3.20 & 2.88 \\ \hline
$f^{95\%}_{\text{min}}$ [TeV] & 0.47 & 0.39 & 3.58 & 3.26 \\ \hline
$f_{\chi^2_\text{min}}$ [TeV] & 1.43 & 0.89 & 13.5 & 8.13 \\ \hline
$\chi^2_{\text{min}}/\text{d.o.f.}$ & 1.048 & 1.041 & 1.049 & 1.043 \\ \hline \hline
$\chi^2_{\text{SM}}/\text{d.o.f.}$ & \multicolumn{4}{|c|}{1.054} \\ \hline
\end{tabular}
\end{center}
\caption{99$\%$ and 95$\%$ \emph{CL} lower bounds on the symmetry
  breaking scale $f$ and $\chi^{2}$ comparison.} 
\label{table:lowerval}
\end{table}

One should notice that there always exists a region in the parameter
space where the measured $\chi^{2}$ is equal or lower than the SM
$\chi^{2}$: however, the minimum $\chi^{2}$ differs only at the 1$\%$
level w.r.t. the SM $\chi^{2}$, so we can conclude that the agreement
of these different $LH$ models with the collected data can be as good
as within the SM, but not significantly better. In particular, for the
$L^{2}H$ and the $SLH$ models the regions of equal or lower $\chi^{2}$
than the SM $\chi^{2}$ shrink to the SM-like decoupling limit. 

The 99$\%$ \emph{CL} lower limits of $f$
can be translated into 99$\%$ \emph{CL} lower limits on the spectrum
of the new heavy particles of the different models, as summarized in
Table $\ref{table:masses}$.
\begin{table}[!ht]
\begin{center}
\begin{tabular}[c]{|c||c|c|} \hline
\textbf{\emph{LHT}} & $m_{\text{min}}$ [GeV], \textbf{Case \emph{A}} & $m_{\text{min}}$ [GeV], \textbf{Case \emph{B}} \\ \hline \hline
$m_{W_{H}}=m_{Z_{H}}$ & 269.6 & 262.2 \\ \hline
$m_{A_{H}}$ & 64.5 & 62.8 \\ \hline
$m_{\Phi}$ & 291.7 & 283.7 \\ \hline
$m_{T_{+}}$ & 553.6 & 537.5 \\ \hline
\end{tabular}
\end{center}

\begin{center}
\begin{tabular}[c]{|c||c|} \hline
\textbf{\emph{L$^2$H}} & $m_{\text{min}}$ [TeV] \\ \hline \hline
$m_{W_{H}}=m_{Z_{H}}$ & 2.13 \\ \hline
$m_{\Phi}$ & 2.30 \\ \hline
$m_{T}$ & 4.50 \\ \hline
\end{tabular}
\hspace{5pt}
\begin{tabular}[c]{|c||c|} \hline
\textbf{\emph{SLH}} & $m_{\text{min}}$ [TeV] \\ \hline \hline
$m_{W_{H}}$ & 1.35 \\ \hline
$m_{Z_{H}}$ & 1.64 \\ \hline
$m_{T}$ & 2.81 \\ \hline
\end{tabular}
\end{center}
\caption{99$\%$ \emph{CL} lower limits on the spectrum
of the new heavy particles.}
\label{table:masses}
\end{table}

The collective symmetry breaking mechanism implemented in each
\emph{LH} model eliminates all 1-loop quadratic divergences in the
Higgs mass squared parameter, where the divergences from the SM
particles are cancelled by quadratically divergent contributions from new
particles with same spin as the respective SM partners. The Higgs mass
squared parameter is thus only logarithmic-divergent at 1-loop. As the
masses of the new particles increase, the difference between the
remaining SM 1-loop contributions and that of the new particles grows,
requiring larger fine tuning of the Higgs mass squared parameter. The
naturalness of the model could therefore be quantified observing by
how much the contributions from the heavy states ($\delta \mu^2$)
exceed the \emph{observed} value of the Higgs mass squared parameter,
as originally proposed in \cite{ArkaniHamed:2002qy}: 
\begin{equation}
	\Delta=\frac{|\delta \mu^2|}{\mu_{\text{obs}}^2}, \qquad \mu_{\text{obs}}^2= \frac{m_h^2}{2}.
	\label{finetun}
\end{equation}
For example, if the new contributions to the Higgs mass squared
parameter exceed $\mu_{\text{obs}}^2$ by a factor of 5, i.e. $\Delta =
5$, one says that the model requires $20\%$ of fine tuning. Clearly,
the lower the value of fine tuning, the worse is the naturalness of
the model. 

The dominant log-divergent contribution to the Higgs mass squared
parameter comes from the top and its heavy partner loops, and is given
for all the three \emph{LH} models we are considering by
\cite{ArkaniHamed:2002qy} 
\begin{equation}
	\delta \mu^2 = -\frac{3 \lambda_t^2 m_T^2}{8 \pi^2} \log{\frac{\Lambda^2}{m_T^2}}
\end{equation}
where $\Lambda=4 \pi f$ is the cut-off of the non-linear sigma model,
$\lambda_t$ is the SM top Yukawa coupling and $m_T$ is the mass of the
heavy top partner as defined in the different models. The thick black
lines on the plots of Fig.~$\ref{chi2fit}$ enclose these regions of
required fine tuning (on the right hand side of the line), and the
level of fine-tuning is also denoted on the plots. 

We can see that the lowest level of fine-tuning is $\sim 10\%$ for both
Case \emph{A} and \emph{B} of \emph{LHT}, while significantly worse
for both \emph{SLH} ($\sim 1\%$) and \emph{L$^2$H} ($\sim 0.1\%$).
Comparing the naturalness of the model, accommodating the 7
and 8 TeV LHC results and the EWPO, to the MSSM~\cite{Berger:2012ec}
shows that only the model with $T$-parity, \emph{LHT}, has less
fine-tuning than the $\sim 1\%$ of the MSSM (in certain regions of
parameter space). This is because the implemented \emph{T}-parity
relieves the constraints from EWPO, allowing a smaller value of the
symmetry breaking scale $f$ and therefore a smaller mass for the
\emph{T}-even top partner. 

\begin{figure}[!ht]
\centering
\includegraphics[width=0.483\textwidth]{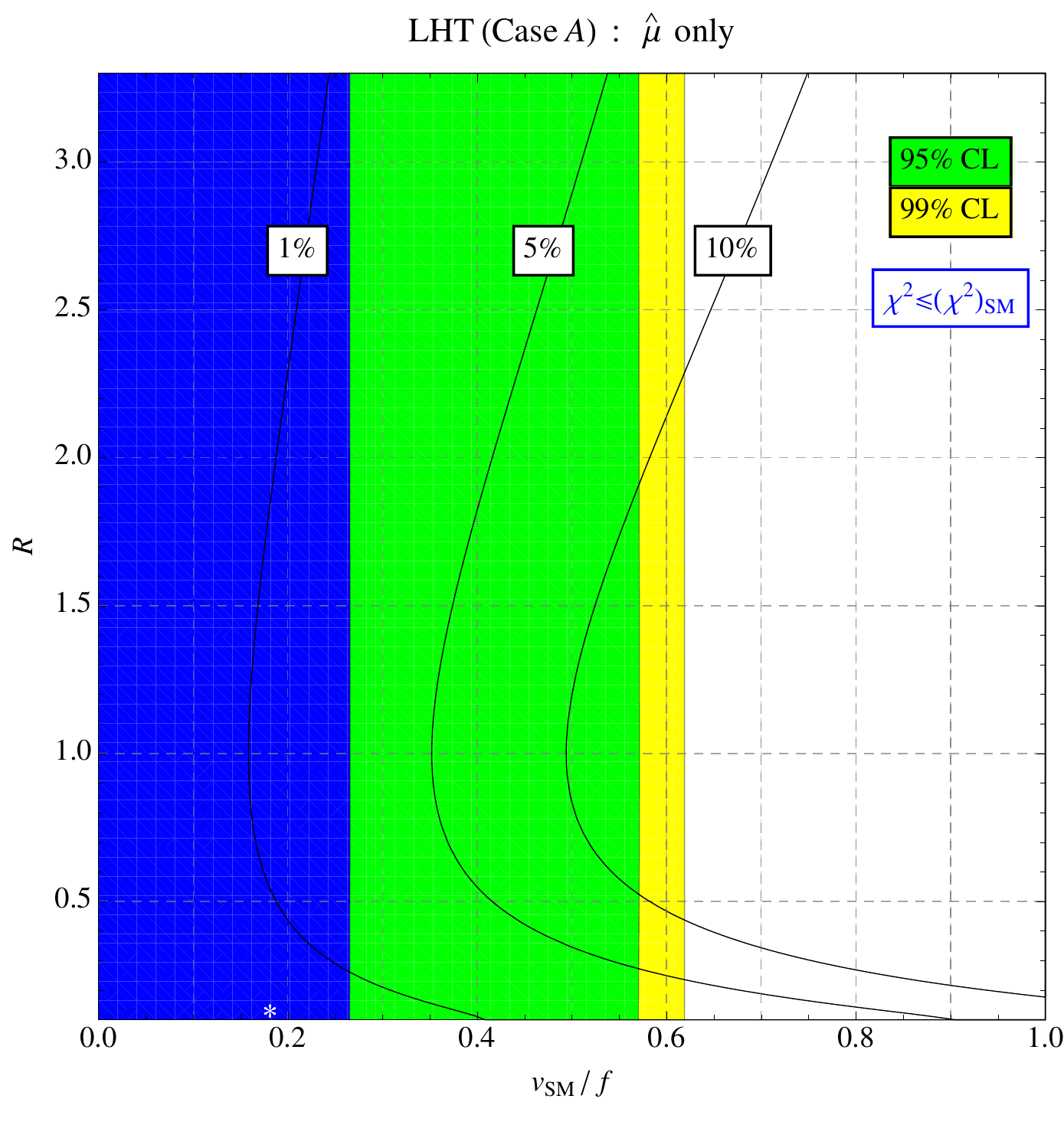} \quad
\includegraphics[width=0.483\textwidth]{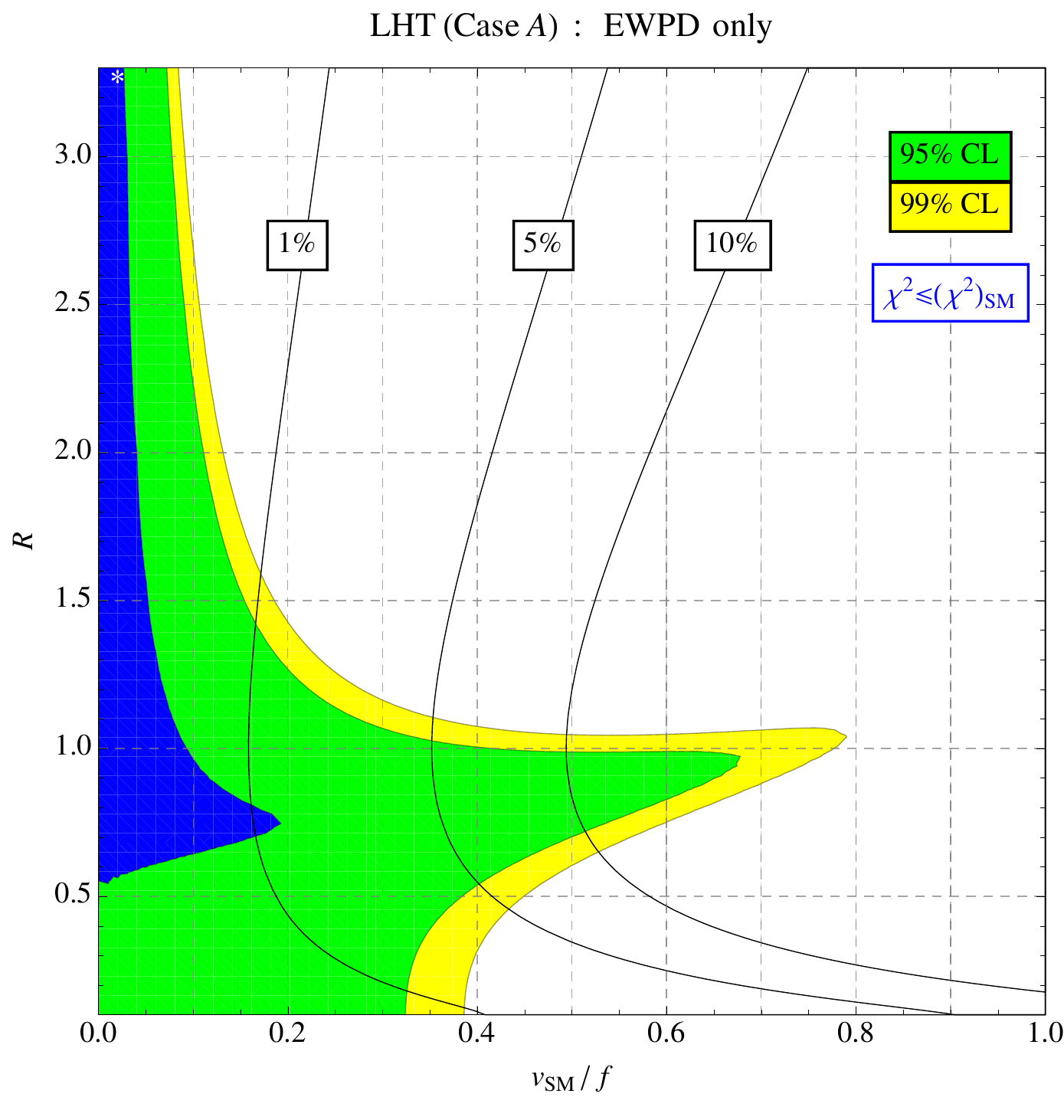}
\caption{Allowed contours at 95$\%$ and 99$\%$ \emph{CL} for
  \emph{LHT} \emph{Case A} considering separately the contributions
  from the Higgs sector only (left) and from EWPO only (right).} 
\label{EWPDvsHiggsLHT}
\end{figure}

It is interesting to consider separately the $\chi^2$ contributions
from the best-fit values and from the EWPO. Considering e.g. the
\emph{LHT} case, the resulting plots are given in
Fig.~\ref{EWPDvsHiggsLHT}~\footnote{We consider only the results for
  Case \emph{A}, as they are compatible 
  with those of Case \emph{B}}.

Clearly, the combined results Fig.~\ref{chi2fit} are mainly driven by
the electroweak data: this should not be surprising, since the
uncertainties on the electroweak observables are much smaller
compared to the uncertainties on the best-fit values of the signal strength
modifiers. This latter statement is therefore true also for the other
\emph{LH} models considered. 

If only the Higgs data are considered, one can notice that there is a
subdominant dependence on the parameter $R$ compared to the ratio
$v_{SM}/f$: this recovers a result already pointed out in
the context of the Higgs Low-Energy Theorem (\emph{LET}) in Composite
Higgs models~\cite{Falkowski:2007hz,Low:2010mr,Furlan:2011uq,Azatov:2011qy,Gillioz:2012se}, namely
that the effective $hgg$ and $h\gamma \gamma$ vertices do not depend
on the details of the heavy fermion sector (in our case on $R$). We
will give the argument in the sequel.

Focusing only on the partial decay widths into two gluons, in the
\emph{LET} approximation the interaction of the Higgs boson with
gluons mediated by loops of colored particles, can be
expressed (at leading order in the expansion of the Higgs
field $h$ around its vacuum expectation value \emph{v} and
considering only the contributions from fermions) by the following 
effective Lagrangian~\cite{Gillioz:2012se} 
\begin{equation}
	\mathcal{L}_{\text{hgg}} = \frac{g_s^2}{48 \pi^2} G_{\mu
          \nu}^a G^{a \, \mu \nu} \frac{h}{v} \left[ \frac{1}{2} v
          \frac{\partial}{\partial v} \log{ \text{det}
            \mathcal{M}^{\dagger} \mathcal{M}_{|_{h=v}} } \right] \; ,
	\label{eff}
\end{equation}
where $\mathcal{M}$ is the fermion mass matrix, including both the
SM-like top and its heavy partner(s). In the narrow width
approximation, the partial width into two gluons normalized to its SM
value is given by the square of the expression in square brackets in
Eq. ($\ref{eff}$), and agrees with our exact result ($\ref{hgg}$) in
the limit of heavy masses running in the loop. 
 
It is a general result \cite{Azatov:2011qy} of Composite Higgs models,
as well as Little Higgs models, that the determinant of the fermion
mass matrix $\mathcal{M}^{\dagger} \mathcal{M}$ is only a function
of the non-linear sigma model expansion parameter $v/f$ and the
details of the heavy fermion sector (i.e. on the masses and couplings
of the fermions), but not separately on $v$: 
\begin{equation}
	\text{det} \mathcal{M}^{\dagger} \mathcal{M}_{|_{h=v}} = F
        \left( \frac{v}{f} \right) \times P \left( \lambda_i, m_i, f
        \right). 
\end{equation}
This factorization clearly makes both Eq.~($\ref{eff}$), and thus also
the partial width into two gluons, independent of the couplings and
masses of the fermions. It is only a function of the non-linear sigma
model  expansion parameter $v/f$. 

Specializing to the \emph{LHT} model case, one can easily see that
this factorization indeed happens by considering the fermion mass
matrix of Eq.~($\ref{matrix}$): 
\begin{eqnarray}
	\text{det} \mathcal{M}^{\dagger} \mathcal{M}_{|_{h=v}} &=&
        \frac{1}{2} \lambda_1^2 \lambda_2^2 f^4 \sin{ \frac{\sqrt{2}
            v}{f} } \nonumber \\ 
	\frac{1}{2} v \frac{\partial}{\partial v} \log{ \text{det}
          \mathcal{M}^{\dagger} \mathcal{M}_{|_{h=v}} } &=& 1-
        \frac{2}{3} \frac{v}{f} 
	\label{collectiveSB}
\end{eqnarray}
making the partial width into two gluons independent of the fermion
couplings (i.e. of $R$), exactly in the \emph{LET} limit and in good
approximation with the exact expression, Eq.~($\ref{hgg}$).  

Analogous statements hold also for the effective coupling of the Higgs
to two photons~\cite{Gillioz:2012se}, so that we conclude that
the partial width into two photons is exactly independent of $R$ in
the \emph{LET} limit, and in good approximation with the exact
expression, Eq.~($\ref{hgaga}$). 

Note that the \emph{L$^2$H} model shares with the
\emph{LHT} model the same form of the top-Yukawa Lagrangian, and
therefore also the fermion mass matrix, cf.~Eqs.~($\ref{topLH}$) and
($\ref{topLHT}$). This allows us to take over the considerations
about the dependence on model parameters for the \emph{L$^2$H} model:
in particular, the results of the Higgs sector will again not depend
on the ratio $R$. Moreover, since the contribution from the heavy
quark loop to the EWPO has been neglected (justified as in
Ref.~\cite{Csaki:2002qg}), we were thus allowed to fix 
the value of $R$ to a reference value ($R$=1) without loss of
generality. 


Another observation from the Higgs-only plot on the left of
Fig.~\ref{EWPDvsHiggsLHT}
is that the region with $v/f \gtrsim 0.62$ is highly disfavored by
the collected data. This is due to the fact that for $f \lesssim
396.2$ GeV, the decay of the $m_h=126$ GeV Higgs boson into two heavy
photons $A_H$ becomes open and dominant, highly reducing all other
branching ratios and therefore the respective predicted signal
strength modifiers $\mu^i$ ($\ref{signalmod}$), clearly in tension
with the observed data, cf. Table~\ref{table:bestfit}.  

An enhancement in the production cross sections could compensate this
reduction in the branching ratios, but this is not the case for the
\emph{LHT} model, since all production modes are reduced w.r.t. their
SM value. The VBF and HS production cross sections are slightly
suppressed because of the suppressed coupling of the Higgs boson with
SM gauge bosons, cf.~Eqs.~($\ref{VBF}$) and ($\ref{HS}$). The GF
production cross section is also suppressed in the whole parameter
space compared to the SM prediction.
This suppression of the GF production in
\emph{LHT} was already pointed 
out in~\cite{Wang:2011rv} in terms of the effective \emph{hgg}
coupling, Eq.~($\ref{hgg}$). Indeed, the top Yukawa coupling is
suppressed by means of the expansion of the non-linear sigma model,
Eq.~($\ref{topyuk}$), and the contribution from the $T_+$ partner (of
opposite sign w.r.t. the top coupling) further suppresses the
\emph{hgg} coupling, as an effect of the collective symmetry breaking
mechanism, as explained before in the context of the \emph{LET}
theorem, cf. Eq. ($\ref{collectiveSB}$). The contribution from the
three \emph{T}-odd partners $u_{1}$ is also negative, as we can see
from Eq.~($\ref{tildetyuk}$). All these contributions cause thus a
suppression in the effective \emph{hgg} coupling ($\ref{hgg}$)
compared to its SM value. 

Considering now the \emph{L$^2$H} model, as for the
\emph{LHT} case the combined result on the lower left of
Fig.~\ref{chi2fit} is mainly driven by the EWPO contribution, and even
more dramatically since \emph{T}-parity eliminates all tree-level
contributions to the oblique parameters from the heavy states and from
the triplet \emph{vev} $v^{\prime}$, which are on the contrary present
in the \emph{L$^2$H} case. 

\begin{figure}[!ht]
\centering
\includegraphics[width=0.483\textwidth]{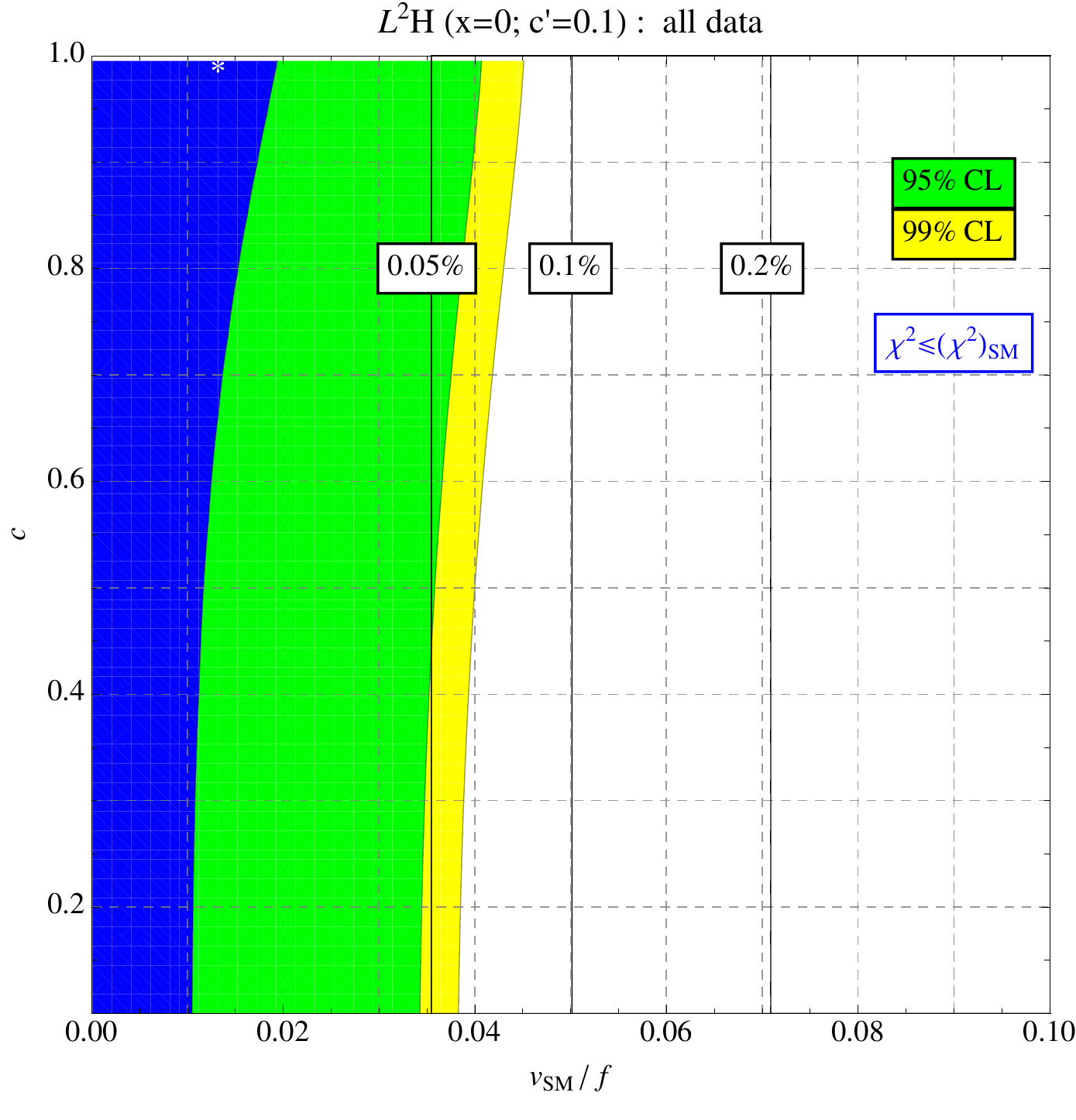} \quad 
\includegraphics[width=0.483\textwidth]{LH_x0_c07.pdf}
\includegraphics[width=0.483\textwidth]{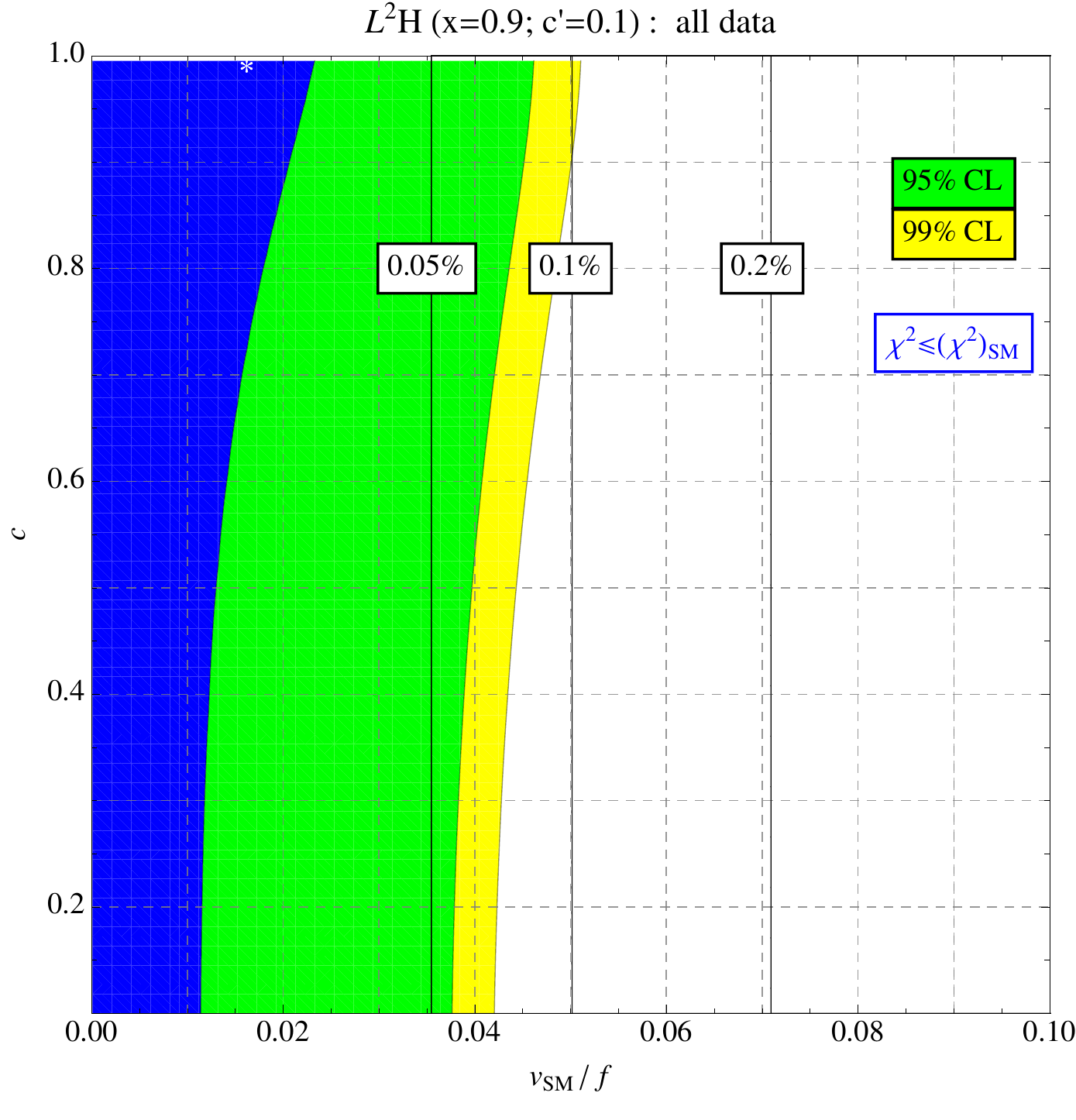} \quad 
\includegraphics[width=0.483\textwidth]{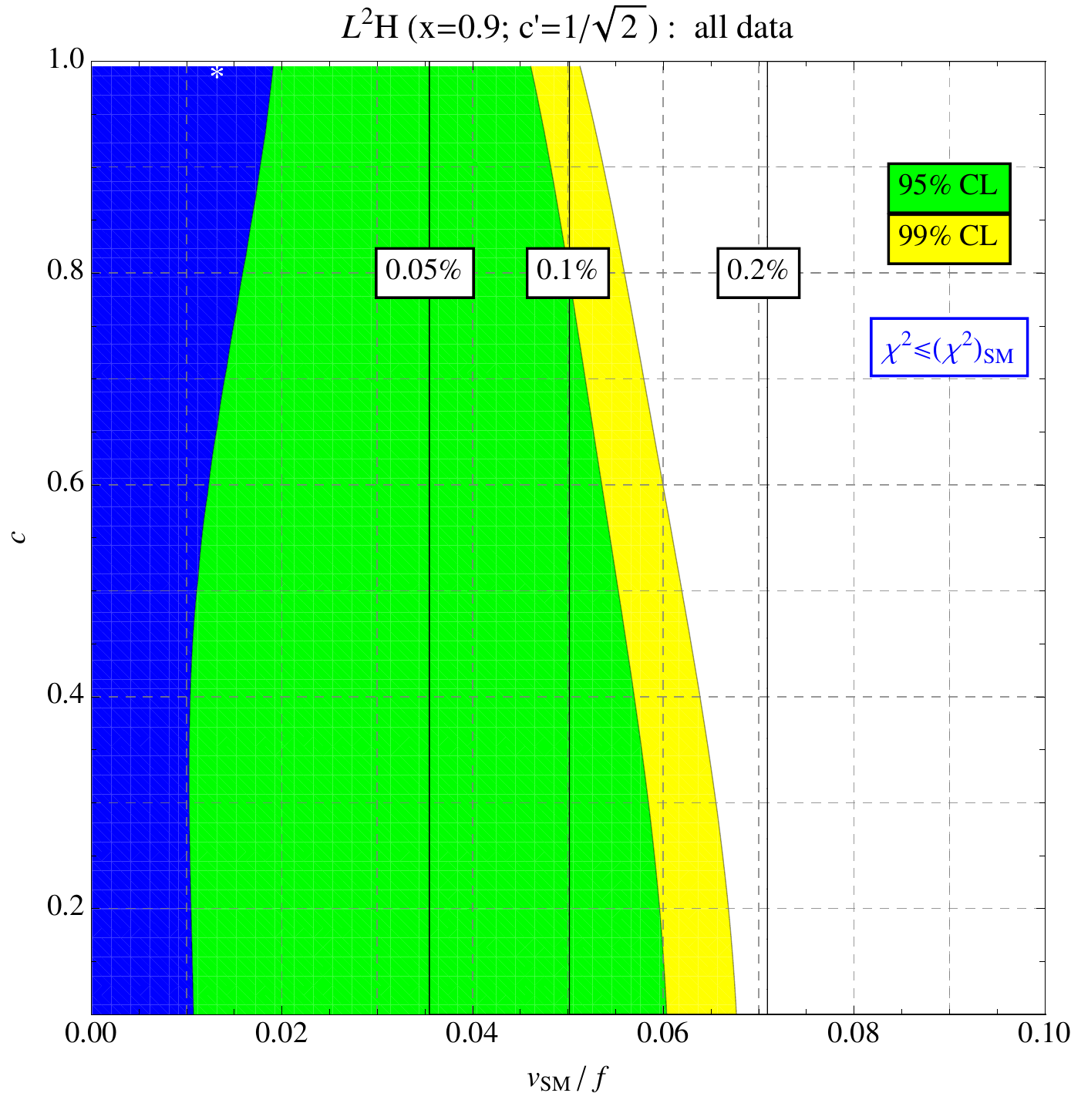}
\caption{\label{fig:params} Allowed contours at 95$\%$ and 99$\%$
  \emph{CL} for \emph{L$^2$H} with different choices of $c^{\prime}$
  and $x$.}  
\end{figure}

In Fig.~\ref{fig:params}, we show the dependence of the results on
different choices of the parameters $c^{\prime}$ and $x$, in
particular with $c^{\prime}= \{ 0.1, 1/\sqrt{2} \}$, i.e.~with minimal
and maximal mixing, respectively, and with $x=\{ 0.0, 0.9 \}$,
i.e.~with vanishing or nearly maximal triplet \emph{vev}.

We can see that there is only a smooth dependence on $c^{\prime}$ and
$x$: the result is driven mainly by the value of the symmetry breaking
scale $f$. Compared to the \emph{LHT} case, the parameter space of the
\emph{L$^2$H} model is indeed highly constrained for lower values of
$v/f$, in particular $v/f \lesssim 0.1$, and this translates into a
higher amount of required fine tuning, of the order $\sim 0.1\%$. 

\begin{figure}[!ht]
\centering
\includegraphics[width=0.483\textwidth]{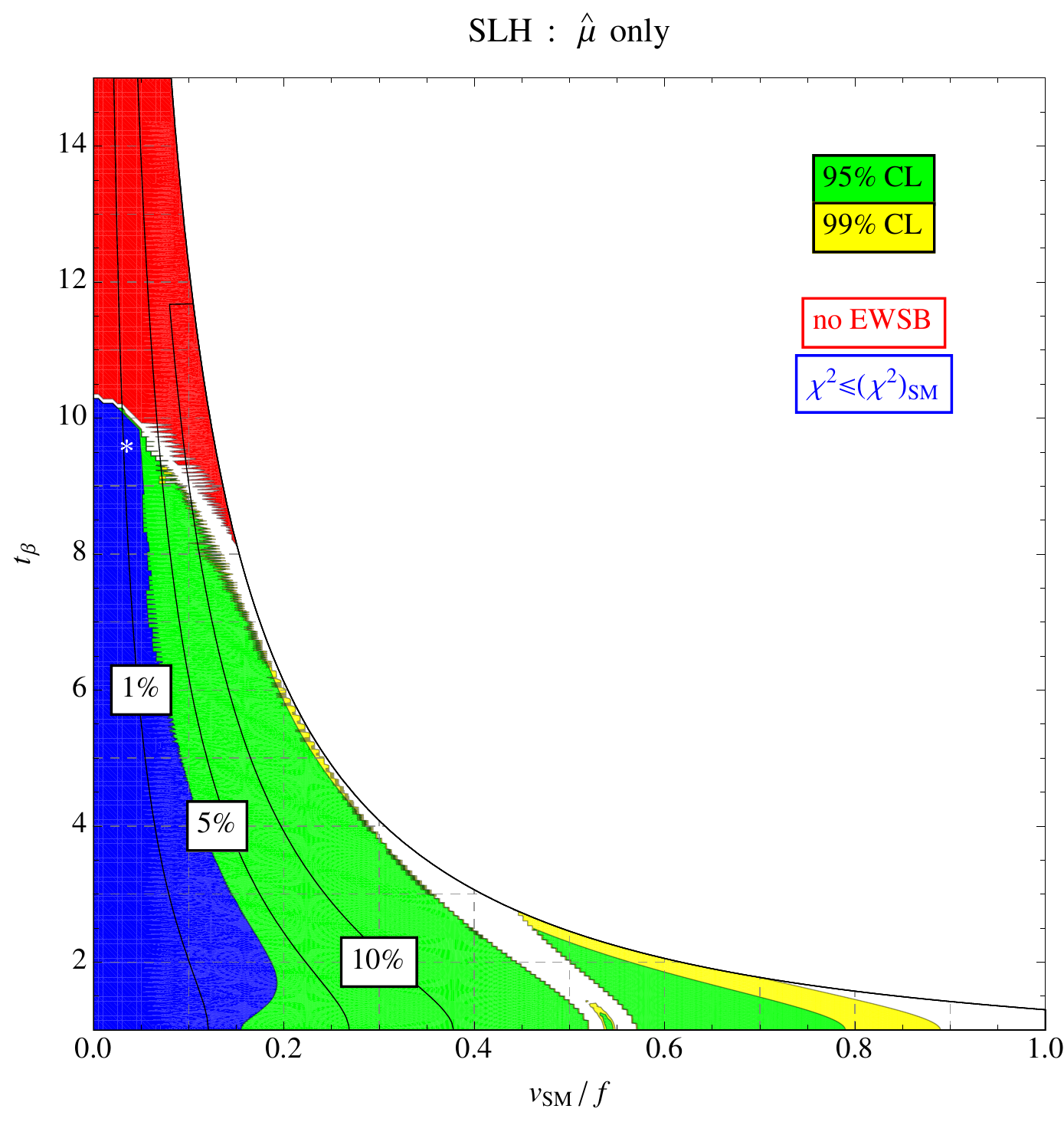} \quad
\includegraphics[width=0.483\textwidth]{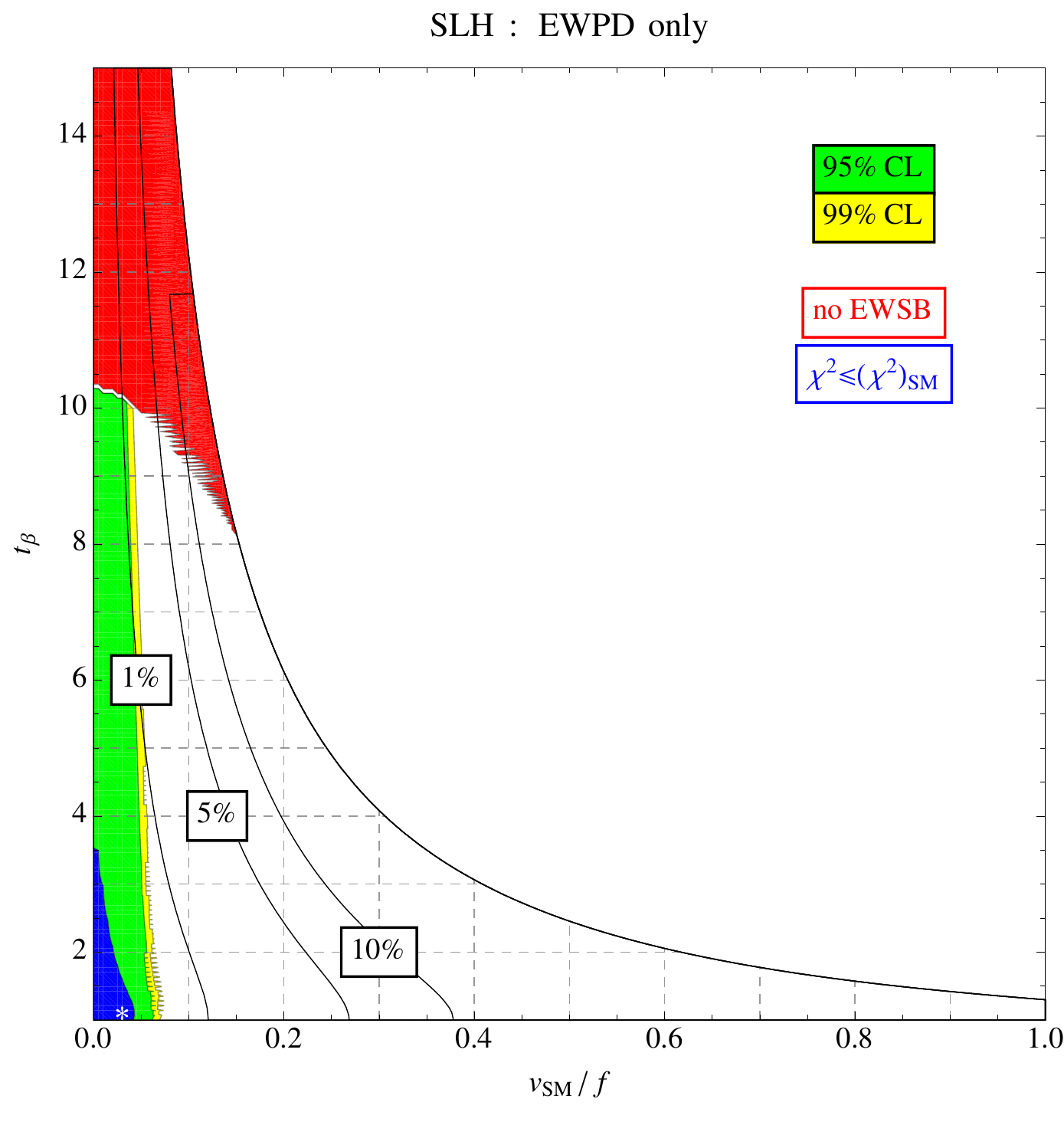}
\caption{\label{fig:l2h} Allowed contours at 95$\%$ and 99$\%$ \emph{CL} for
  \emph{SLH} considering separately the contributions from the Higgs
  sector only (left) and from EWPO only (right).} 
\label{EWPDvsHiggsSLH}
\end{figure}

Not surprisingly, also the results for the \emph{SLH} are driven by
the EWPO constraints, as can be seen from Fig.~\ref{fig:l2h}.

Nearly the whole parameter space is indeed still compatible at 99$\%$
\emph{CL} with the Higgs-sector data: only new data with increasing
luminosity and reduced uncertainties on the best-fit values of the
signal-strength modifiers could give us more stringent information.  

\begin{figure}[!ht]
\centering
\includegraphics[width=0.7\textwidth]{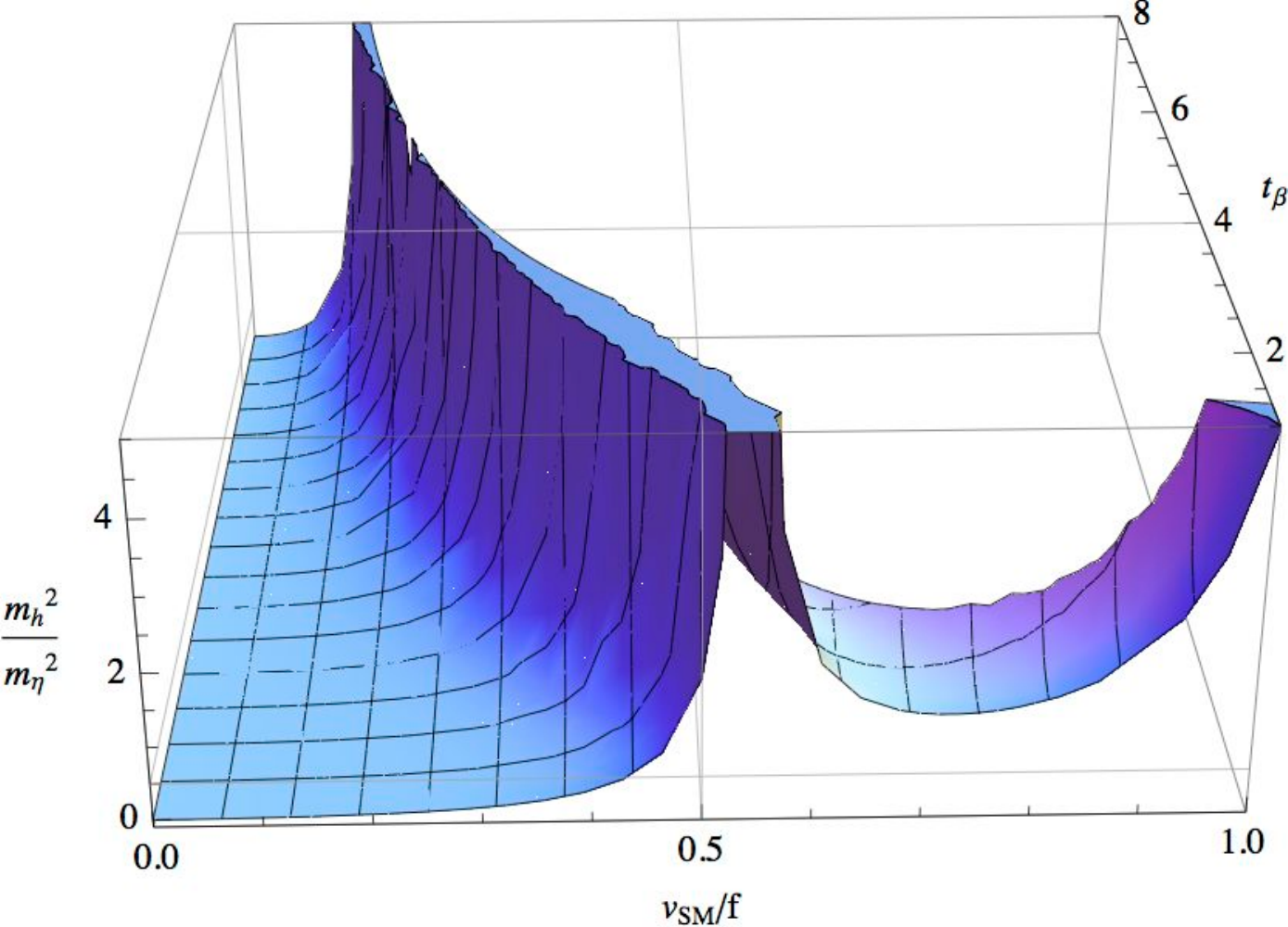}
\caption{\label{fig:mass} Higgs mass versus pseudo-scalar $\eta$ mass
  in \emph{SLH}.}  
\end{figure}

It is to be noted that in the Higgs-data plot there is however an
excluded central band up to $v_{SM}/f \sim 0.6$: in this region, the
decays of the Higgs involving the pseudo-scalar $\eta$ are indeed open
and dominant, cf. Eq. ($\ref{hetaeta}$) and ($\ref{hZeta}$), highly
reducing all other SM-like branching ratios, in the same way as for
the decay of the Higgs into a pair of heavy photons $A_H$ discussed in
the \emph{LHT} case. Indeed if we plot the ratio of the Higgs mass
w.r.t. the mass of the pseudo-scalar $\eta$, we can identify the
excluded regions in the left plot of Fig.~$\ref{EWPDvsHiggsSLH}$ with
the regions where the tree level decays involving the pseudo-scalar are  
kinematically accessible, cf.~Fig.~\ref{fig:mass}.


\section{Conclusions}
\label{sec:conclusions}

In this paper, we have investigated the parameter space of the most 
prominent member of Little Higgs models, the Littlest Higgs model with
and without $T$-parity from the class of Product Group Models, as well
as the Simplest Little Higgs from the class of Simple Group Models, in
the light of all present collider data as of the end of the year
2012. We included all published discovery and search channels for
the Higgs boson from both LHC collaborations, ATLAS and CMS, together
with the electroweak precision observables. The latter have been
mostly updated by the results on the $W$ mass from the Tevatron
experiments recently. 

Our results show that the experimental data from LHC are not yet
precise enough to compete with the electroweak precision data, and we
have to wait for un update from the collaborations with higher
luminosity. For both the Littlest Higgs and the Simplest Little Higgs
model without $T$-parity, EWPO force the Little Higgs scale $f$ to be
of the order of 2-4 TeV in order to be compatible with the precision
electroweak data. Both models hence show a bit worse fine-tuning than
the so-called natural pMSSM. On the other hand, $T$-parity does the
job for which it has been invented, namely to reduce this amount of
fine-tuning. The scale in the case of the Littlest Higgs model with
$T$-parity is only constrained to be above 700-1200 GeV, and in most
cases, the new $T$-odd particles can still be well below the TeV
scale. The fine-tuning in the Higgs sector is less by a factor of two
to five compared to the natural pMSSM. 

We note further, that we did not include searches for exotic particles
like additional gauge bosons or heavy vector-like quarks in our fits. 
This has been partially done
elsewhere~\cite{Berger:2012ec,Perelstein:2011ds,Godfrey:2012tf}, and
on the other hand, these 
searches will not be reaching enough sensitivity before the start
of the 14 TeV run to become truly compatible. 


\vskip0.8cm
\noindent{\large \bf Acknowledgments} 
\vskip0.3cm
We are grateful to M.~Asano, J.~Galloway, M.~Perelstein, K.~Sakurai and
A.~Weiler for valuable remarks and discussions. M.T. acknowledges
financial support by the Sonderforschungsbereich 676 ``From Strings to
Particles'' of the German Science Foundation (DFG). 


\newpage

\appendix

\section{Higgs Boson Partial Widths and Production Cross Sections}
\label{app:higgs}

For tree-level decays of the Higgs, the partial widths get a
correction at lowest order via the corresponding modified couplings 
\cite{Wang:2011rv}: 
\begin{eqnarray}
	\Gamma(h \rightarrow VV) &=& \Gamma(h \rightarrow VV)_{SM}
        \Bigg( \frac{g_{hVV}}{g_{hVV}^{SM}} \Bigg)^{2} \qquad \text{$V
          \equiv W,Z$} \nonumber \\ 
	\Gamma (h \rightarrow \bar{f} f) &=& \Gamma(h \rightarrow
        \bar{f} f)_{SM} \Bigg( \frac{g_{hff}}{g_{hff}^{SM}} \Bigg)^{2}
        \qquad \text{$f \equiv c,b,\mu,\tau$} 
\end{eqnarray}
with the different couplings and masses defined as in the previous sections.

There are also some new tree-level decay channels which are special to
the different Little Higgs models, and which have to be taken into
account if kinematically accessible. In particular, defining 
\begin{equation}
	x_{i} = \frac{4 \, m_{i}^{2}}{m_{h}^{2}}
\end{equation}
in \emph{LHT} the Higgs field could decay into two heavy photons $A_H$, basically an invisible decay, with partial width \cite{Wang:2011rv}
\begin{equation}
	\Gamma (h \rightarrow A_{H} A_{H}) = \frac{g_{h A_{H} A_{H}}^{2} \, m_{h}^{3}}{128 \, \pi \, m_{A_{H}}^{4}} \sqrt{1- x_{A_{H}}} \left( 1- x_{A_{H}} + \frac{3}{4} x_{A_{H}}^{2} \right) \qquad \text{if $x_{A_{H}}<1$}
	\label{hAHAH}
\end{equation}
while in \emph{SLH} two new decay channels involving the pseudo-scalar $\eta$ are possibly open \cite{Cheung:2006nk}
\begin{equation}
	\Gamma (h \rightarrow \eta \eta) = \frac{m_{\eta}^{4}}{8 \, \pi \, v^{2} \, m_{h}} \sqrt{1-x_{\eta}} \qquad \text{if $x_{\eta} <1$}
	\label{hetaeta}
\end{equation}
\begin{equation}
	\Gamma (h \rightarrow Z \eta) = \frac{m_{h}^{3}}{32 \, \pi \, f^{2}} \left( \frac{\tb^{2}-1}{\tb} \right)^{2} \lambda^{3/2} \left( 1, \frac{m_{Z}^{2}}{m_{h}^{2}}, \frac{m_{\eta}^{2}}{m_{h}^{2}} \right)
	\label{hZeta}
\end{equation}
with $\lambda (1,x,y)= \left( 1-x-y \right)^{2} -4 xy$. 
\vspace{10pt}

Defining the functions \cite{Gunion:1989we}
\begin{eqnarray}
	F_{0} (x) &=& x \left[ 1- x f(x) \right] \nonumber \\
	F_{1/2} (x) &=& -2x \left[ 1+(1-x) f(x) \right] \nonumber \\
	F_{1}(x) &=& 2 + 3x + 3x (2-x) f(x) \nonumber \\
	f(x) &=& \begin{cases}
		\left[ \sin^{-1}{\left( \frac{1}{\sqrt{x}} \right)} \right]^{2} & \text{for } x \geq 1 \\
		- \frac{1}{4} \left[ \log \left( \frac{1+
                      \sqrt{1-x}}{1- \sqrt{1-x}}  \right) - i \pi
                \right]^{2} & \text{for } x < 1
              \end{cases} \nonumber
\end{eqnarray}
the general expression of the partial widths for the one-loop decays
of the Higgs boson into two gluons or two photons are 
given by 
\begin{equation}
	\Gamma(h \rightarrow gg) = \frac{ \alpha_{s}^{2} \, m_{h}^{3}
        }{32 \, \pi^{3} \, v^2} \Big| \sum_{f, \, \text{col}}
        -\frac{1}{2} \, F_{1/2}(x_f) \, y_f \Big|^{2} \qquad ,
	\label{hgg}
\end{equation}
where the sum is extended over all colored fermionic
particles of the spectrum which have a non-negligible coupling $y_f$
to the Higgs boson (in contrast to SUSY there are no colored
scalars in Little Higgs models), and by 
\begin{equation} 
	\Gamma(h \rightarrow \gamma \gamma) = \frac{ \alpha^{2} \,
          m_{h}^{3}}{256 \, \pi^{3} \, v^2} \Big| \sum_{f, \,
          \text{ch}} \frac{4}{3} \, F_{1/2}(x_f) \, y_f + \sum_{v, \,
          \text{ch}} F_{1} (x_v) \, y_v + \sum_{s, \, \text{ch}}
        F_{0}(x_s) \, y_s \Big|^{2} 
	\label{hgaga}
\end{equation}
respectively. Here, the different sums run over all
electrically charged fermionic (\emph{f}), vector (\emph{v}) and
scalar (\emph{s}) particles of the spectrum which have a
non-negligible coupling $y_{f,v,s}$ to the Higgs boson. 
\vspace{10pt}

At the LHC, the main production channel for the Higgs is the Gluon Fusion (GF):
\begin{figure}[!ht]
\centering
\includegraphics[width=0.3\textwidth]{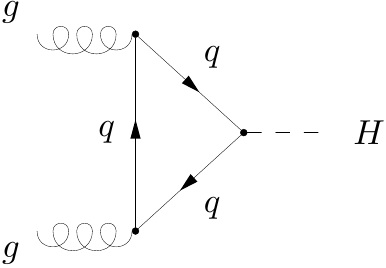}
\end{figure}

The hadronic GF cross section is given by the usual convolution \cite{Gunion:1989we}
\begin{equation}
	\sigma \left( pp \rightarrow h \right) = \int_{\tau}^{1} \frac{dx}{x} g \left( x,\mu_{F}^{2} \right) g \left( \frac{\tau}{x},\mu_{F}^{2} \right) \hat{\sigma} \left( gg \rightarrow h \right) \, \tau
\end{equation}
where $\tau= m_{h}^{2}/s$, with $s$ the total hadronic c.m.~energy
squared, $g \left( x, \mu_{F}^{2} \right)$ is the parton distribution
function of the gluon at the factorization scale $\mu_F^2$, and  
\begin{equation}
	\hat{\sigma} \left( gg \rightarrow h \right)= \frac{\pi^{2}}{8 \, m_{h}^{3}} \, \Gamma \left( h \rightarrow gg \right)
	\label{narrow}
\end{equation}
is the partonic cross section in the narrow-width approximation. 

Using Eq. ($\ref{narrow}$) we can thus approximate the Little Higgs prediction for the GF cross section as a rescaling of the SM prediction:
\begin{equation}
	\sigma(pp \rightarrow h)_{\text{LH}} \sim \frac{\Gamma \left(
            h \rightarrow gg \right)_{LH}}{\Gamma \left( h \rightarrow
            gg \right)_{SM}} \cdot \, \sigma(pp \rightarrow
        h)_{\text{SM}}. 
	\label{gfusion}
\end{equation}

The second most important channel for Higgs production at the LHC is
the vector-boson Fusion (VBF). For the mass of the Higgs we are
considering, the SM VBF cross section is smaller than the
GF cross section by about an order of magnitude, but it could
be important for some Higgs decay channels because of its distinctive
kinematic signature of forward jets with high transverse momentum. 

To calculate the Little Higgs prediction of the VBF cross section we
have not included the contributions from the heavy gauge bosons (or
any other heavy particles) as they are more difficult to be produced, and
therefore we can safely neglect their contribution to the VBF
cross section. We are left therefore with only the contributions from
the light quarks ($u, \ldots ,b$) and from the $Z$ and $W^{\pm}$ gauge
bosons.  

Neglecting for simplicity possible $\mathcal{O} \left( v/f \right)$
corrections in the charged- and neutral-current couplings of the light
quarks with the SM gauge bosons, at tree-level the Little Higgs VBF
cross section is then given by its SM value rescaled with the
appropriate (and model dependent) Higgs-gauge bosons coupling squared
$a^2 \equiv ( g_{hVV}/g_{hVV}^{SM} )^2$. Hence, we can factorize the
rescaling factor out of the amplitude 
\begin{figure}[!ht]
\centering
\includegraphics[width=0.6\textwidth]{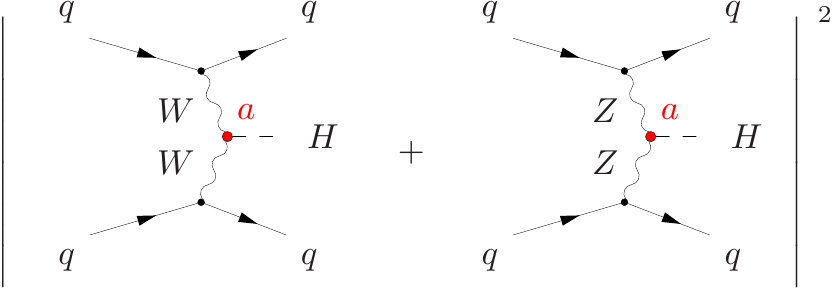}
\end{figure}
\begin{equation}
	\Rightarrow \quad \sigma(qq \rightarrow qqh)_{\text{LH}} =
        \Bigg( \frac{g_{hVV}}{g_{hVV}^{SM}} \Bigg)^2 \cdot \,
        \sigma(qq \rightarrow qqh)_{\text{SM}}. 
	\label{VBF}
\end{equation}
Even if the latter equation is a tree-level result, we have used the
value of the SM VBF cross section recommended in
\cite{Dittmaier:2011ti}, obtained at higher perturbative orders, to
obtain the Little Higgs prediction through Eq. ($\ref{VBF}$). 

One should notice that in order to evaluate the VBF cross section in
the \emph{SLH} model, we neglected also the custodial-symmetry
violating shift in the $Z$-mass ($\ref{ghVVSLH}$), which is
parametrically smaller than the other $\mathcal{O} \left( v^2/f^2
\right)$ contribution, so that both $hWW$ and $hZZ$ vertices could
share the same rescaling factor. 
\vspace{10pt}

The Higgs-Strahlung production (HS) has a cross section which
is about one to two orders of magnitude smaller than the GF cross
section for Higgs masses $<200$ GeV, but it is important for certain
Higgs decay channels because of the possibility of tagging the
associated vector boson in leptonic decays. In particular one can
distinguish between the production of the Higgs associated with the
charged $W^{\pm}$ or with the neutral $Z$.

Using the same arguments as for the VBF case, we can conclude that
at tree-level the correction to the HS cross section is again
proportional to the square of the respective modified Higgs-gauge
bosons coupling 
\begin{figure}[!ht]
\centering
\includegraphics[width=0.9\textwidth]{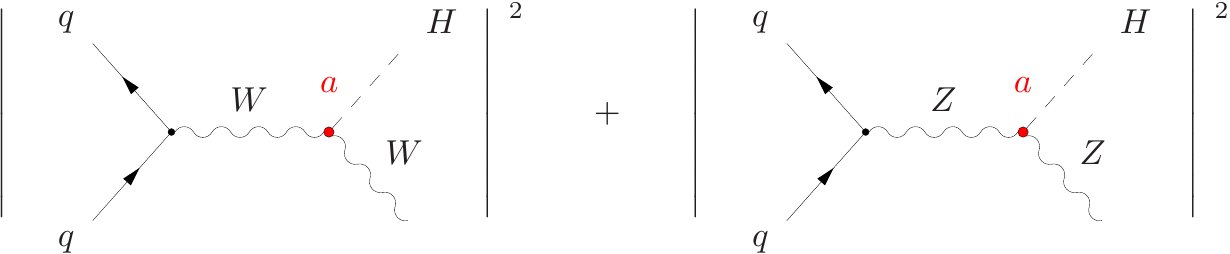}
\end{figure}
\begin{equation}
	\Rightarrow \quad \sigma(qq \rightarrow Vh)_{\text{LH}} =
        \Bigg( \frac{g_{hVV}}{g_{hVV}^{SM}} \Bigg)^2 \cdot \,
        \sigma(qq \rightarrow Vh)_{\text{SM}}. 
	\label{HS}
\end{equation}
As before, even if the latter equation is a tree-level result, we have
used the value of the SM HS cross sections recommended in
\cite{Dittmaier:2011ti}, obtained at higher perturbative orders, to
obtain the Little Higgs prediction through Eq. ($\ref{HS}$). 


\section{General structure of Electroweak Precision Observables}
\label{app:ewpo}

Using the notation of Ref.~\cite{Burgess:1993vc}, one can parametrize
the change of the charged- and neutral-current couplings of the
SM-like gauge bosons due to the presence of new-physics as follows 
\begin{eqnarray}
	\mathcal{L}_{cc} &\supset& - \frac{g}{\sqrt{2}} \sum_{i,j}
        \bar{f}_i \gamma^{\mu} \left( (h_L+\delta \tilde{h}_L)P_L +
          (h_R+\delta \tilde{h}_R)P_R \right) f_j W_{\mu} \nonumber \\ 
	\mathcal{L}_{nc} &\supset& -\frac{g}{\tilde{c}_W} \sum_i
        \bar{f}_i \gamma^{\mu} \Big( (g_L+\delta \tilde{g}_L)P_L +
        (g_R+\delta \tilde{g}_R)P_R \Big) f_i Z_{\mu}  \qquad ,
\end{eqnarray}
where $g_{L,R},h_{L,R}$ are normalized such that $g_L= I_3 - Q \cdot
\tilde{s}_W^2$, $g_R=-Q \cdot \tilde{s}_W^2$, $h_L = V_{ij}$, $h_R =
0$. $P_{L,R}$ are the chiral projectors, and $\tilde{c}_W =
g/\sqrt{g^2 + g^{\prime \, 2}}$ (and similarly $\tilde{s}_W$) are the
\emph{bare} couplings as in the Standard Model\footnote{When
  evaluating electroweak observables, one should notice that
  the bare couplings $\tilde{c}_W,\tilde{s}_W,\tilde{e}$ can also
  get corrections from non-zero oblique parameters \emph{S,T,U}.}. 

In Ref.~\cite{Burgess:1993vc}, the authors presented a parametrization
of 21 different EWPO, both at low energies and at the $Z$ pole, in terms
of the oblique parameters \emph{S,T,U} and of the coefficients $\delta
\tilde{g}_{L,R},\delta \tilde{h}_{L,R}$ defined as before. We refer to
the original reference for the explicit expressions. 

In the $L^2H$ model, Ref. \cite{Csaki:2002qg} already provides the
explicit expressions of the 21 EWPO in terms of the free parameters of
the model. For the \emph{LHT} and \emph{SLH} models, we calculated
the different contributions of these models to the oblique
parameters \emph{S,T,U} and to the coefficients $\delta
\tilde{g}_{L,R},\delta \tilde{h}_{L,R}$, in order to reconstruct the
different contributions to the EWPO, which have then been included in
our $\chi^2$ analysis.

\end{document}